\def\gsim{\;\lower4pt\hbox{${\buildrel\displaystyle >\over\sim}$}\,}
\def\lsim{\;\lower4pt\hbox{${\buildrel\displaystyle <\over\sim}$}\,}

\documentclass[usenatbib]{mn2e}
\usepackage{graphicx}

\newcommand\E[1]{\times10^{#1}}
\newcommand\un[1]{{\,\rm #1}}
\newcommand\rs[1]{_\mathrm{#1}}
\newcommand\g{$\gamma$}
\newcommand\ApJ{ApJ}

\title[Constraints on SN~1006]{Observational constraints on the modeling of SN~1006}
\author[Petruk O. et al.]{O.~Petruk$^{1,2}$, V.~Beshley$^{1}$, 
F.~Bocchino$^{3}$, M.~Miceli$^{3}$, S.~Orlando$^{3}$\\
$^{1}$Institute for Applied Problems in Mechanics and Mathematics, Naukova St.\ 3-b,
   79060 Lviv, Ukraine\\
$^{2}$Astronomical Observatory, National University, Kyryla and Methodia St.\ 8, 79008 Lviv, Ukraine\\
$^{3}$INAF — Osservatorio Astronomico di Palermo, Piazza del Parlamento 1, 90134 Palermo, Italy\\
}

\begin{document}

\date{Accepted .... Received ...; in original form ...}

\pagerange{\pageref{firstpage}--\pageref{lastpage}} \pubyear{2008}

\maketitle

\label{firstpage}

\begin{abstract}
Experimental spectra and images of the supernova remnant SN~1006 have
been reported for radio, X-ray and TeV gamma-ray bands. Several
comparisons between models and observations have been discussed in the
literature, showing that the broad-band spectrum from the whole remnant
as well as a sharpest radial profile of the X-ray brightness can be
both fitted by adopting a model of SN~1006 which strongly depends on
the non-linear effects of the accelerated cosmic rays; these models
predict post-shock magnetic field (MF) strengths of the order of
$150\un{\mu G}$. Here we present a new way to compare models
and observations, in order to put constraints on the physical
parameters and mechanisms governing the remnant. In particular,
we show that a simple model based on the classic MHD and
cosmic rays acceleration theories (hereafter the `classic' model)
allows us to investigate the spatially distributed characteristics
of SN~1006 and to put observational constraints on the kinetics and
MF. Our method includes modelling and comparison of the azimuthal and
radial profiles of the surface brightness in radio, hard X-rays and
TeV \g-rays as well as the azimuthal variations of the electron
maximum energy. In addition, this simple model also provides good
fits to the radio-to-gamma-ray spectrum of SN~1006. We find
that our best-fit model predicts an effective MF strength inside SN~1006 of
$32\un{\mu G}$, in good agreement with the `leptonic' model suggested by
the HESS Collaboration (2010).
Finally, some difficulties in both the
classic and the non-linear models are discussed. A number of evidences
about non-uniformity of MF around SN~1006 are noted.
\end{abstract}

\begin{keywords}
{ISM: supernova remnants -- individual:SN~1006 -- ISM: cosmic rays
-- radiation mechanisms: non-thermal -- acceleration of particles 
}
\end{keywords}

\section{Introduction}

The supernova remnant (SNR) SN~1006 is one of the most
interesting objects for studies of Galactic cosmic rays. It is quite
symmetrical with a rather simple bilateral morphology in radio
\citep[e.g.][]{pet-SN1006mf}, nonthermal X-rays \citep[e.g][]{SN1006Marco}
and TeV \g-rays \citep{HESS-SN1006-2010}. Its prominent feature is
the positional coincidence of the two bright nonthermal limbs in all
these bands, including TeV \g-rays as demonstrated by recent results
of \citet{HESS-SN1006-2010}.

Current investigations of SNRs with TeV \g-ray emission demonstrate an
ambiguity in the explanation of the nature of TeV \g-rays. Namely,
the broad-band (radio-to-\g-rays) spectrum of these SNRs can be
fitted by assuming the TeV radiation either as leptonic or as hadronic in
origin \citep[e.g. RX J1713.7-3946:][]{RX1713aha2006,RX1713Ber-Volk-06}.

The question of the origin of TeV \g-rays is closely related to the
problem of the presence and the role of non-linear effects of cosmic
rays acceleration by the forward shock. One of the key parameter
distinguishing between these two possibilities is the strength
(and thus the nature) of the post-shock magnetic field. The
classical picture considers only the compression of the
typical interstellar magnetic field (ISMF) $B\rs{o}\sim 3\un{\mu G}$
to downstream values of the order of tens $\mu$G. Models
including non-linear acceleration (NLA) predict that the ISMF is
first amplified upstream due to the back reaction of accelerated protons
to $B\rs{o}\sim 30\un{\mu}$G and then compressed above hundred
$\mu$G. In the former case the inverse-Compton (IC) emission of electrons
would be responsible for most of the TeV \g-rays, in the latter case
the proton-origin TeV \g-ray radiation is expected to be dominant.

The spectrum of SN~1006 may be explained in these two scenarios.
One limiting possibility (we call it `extreme NLA model'), namely the
case of ISMF amplified and compressed to $B\rs{s}\approx 150\un{\mu G}$
is considered in details by \citet{SN1006Ber-Ksen-Volk-09}. The model
successfully fits the broadband nonthermal spectrum from SN~1006 and the
sharpest radial profile of the X-ray brightness. TeV \g-rays are shown
to be produced in both the inverse-Compton mechanism and the pion-decay
one, the latter is dominant.

Here, we present a new method to compare models and observations. In
particular, we investigate the origin of the patterns of nonthermal
images in radio, X-rays and \g-rays. At present time, this can be
done only by using the classic MHD and particle acceleration
theories. Therefore, the questions behind the present paper are: may a
classical model explain the radio-to-TeV-\g-ray observations of SN~1006
and can one put observational constraints on some properties of the
particle kinetics and/or on the MF?

In this work, we introduce a ``classic'' model describing SN~1006
and compare the spatial distribution of surface brightness derived
from the model with those from observations in different wavelength
bands. The comparison will allow us to put some constraints on the
parameters of the model, thus deriving some hints on the physical
mechanisms governing the cosmic rays acceleration in SN~1006. In the
following, the section order is determined by the order of parameters
determination: the azimuthal and radial profiles in the radio band are
analysed in Sect.~\ref{sn1006cp:radio_section}; the variation of the break
frequency in Sect.~\ref{EmaxTheta0}; the broadband spectrum of SN~1006 is
calculated in Sect.~\ref{SN1006:vol-sp} to check the consistency of our
model and to determine the average MF; the X-ray and \g-ray brightness
are investigated in Sect.~\ref{sn1006cp:xray_section}. Finally, we draw
our conclusions in Sect.~\ref{sn1006cp:conclusion_section}.

\begin{figure*}
\centering
\includegraphics[width=17truecm]{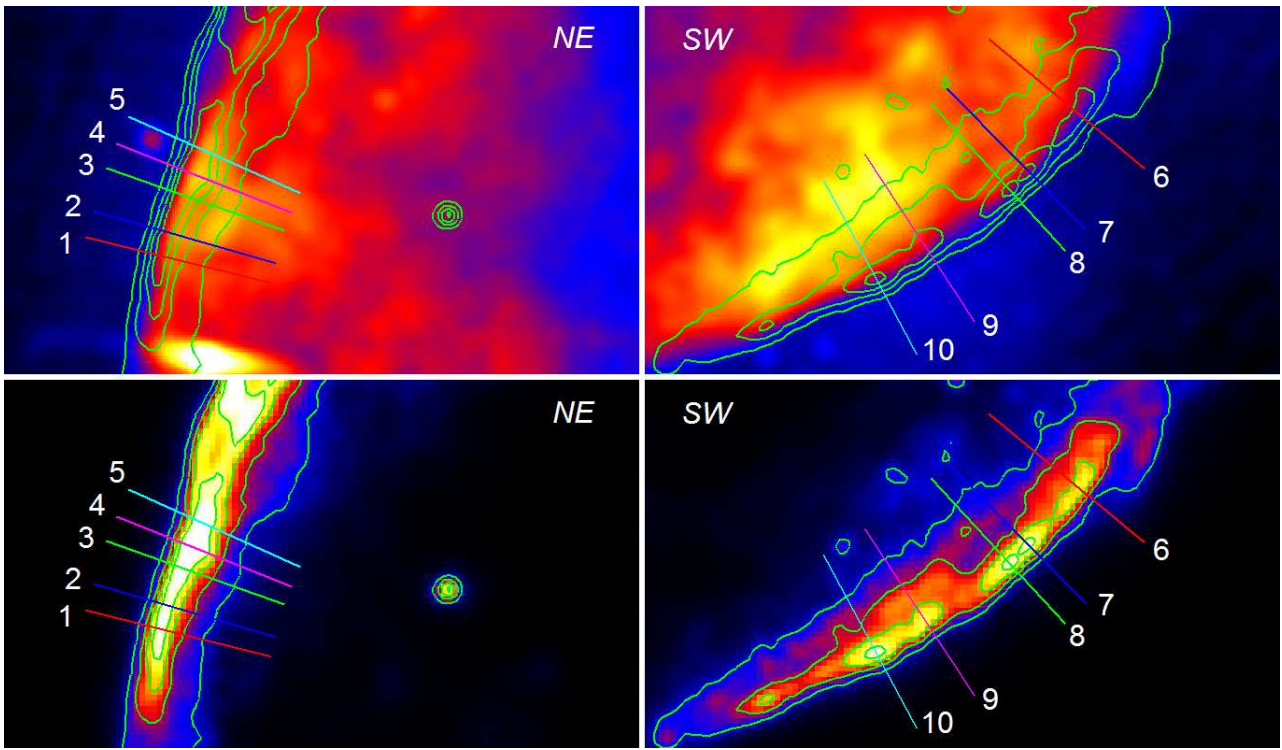}
\caption{NE and SW limbs of SN~1006 in radio at $\lambda\sim20\un{cm}$
(top panels) and X-rays with energy 2-4.5 keV  (bottom panels)
\citep{pet-SN1006mf,SN1006Marco}.  The maximum value of brightness is
100 times the minimum one, in the radio and X-ray images.  Radio image
is smoothed with Gaussian with $\sigma=0.4'$ to lower fluctuations.
Color straight lines mark the regions used for extraction of the
radial profiles of brightness; length of regions shown is from $0.8R$
to $1.1R$. Green lines represent X-ray contours, linearly spaced.}
\label{SN1006cp:fig-a}
\end{figure*}

\section{Constraints from radio maps}
\label{sn1006cp:radio_section}

We consider an SNR expanding through a uniform ISM and
uniform ISMF, in the adiabatic stage of its evolution. The
Sedov solution is therefore appropriate to describe the
hydrodynamics of the system. We consider ideal gas with the
adiabatic index $\gamma=5/3$. The MF evolution is treated in the
classic framework, without non-linear amplification. Its strength
decreases downstream\footnote{At the shock front the
perpendicular component of the MF is enhanced by a factor 4; then the
parallel and perpendicular components evolve independently downstream
of the shock, following the MF flux conservation and flux freezing
conditions respectively \citep{Reyn-98}.} far away from the
shoch front and is modelled following \citet{Reyn-98}; the
role of the ejecta is considered to be negligible. The
classic (unmodified) shock creates the energy spectrum of relativistic
electrons in the form $N(E)dE=KE^{-s}\exp(-E/E\rs{max})dE$ with
the spectral index $s$, normalization $K$ and maximum energy of
accelerated electrons $E\rs{max}$. Their dependences on obliquity
are denoted as $K\rs{s}(\Theta\rs{o})=K\rs{s\|}f\rs{K}(\Theta\rs{o})$,
$E\rs{max}(\Theta\rs{o})=E\rs{max\|}f\rs{E}(\Theta\rs{o})$ where the
symbol `$\|$' marks values at the parallel shock. The downstream
evolution of the electron spectrum is modelled as in \citet{Reyn-98}.

\subsection{Azimuthal profiles}
\label{sn1006cp:radio_azimuth}

The symmetrical bright limbs in SN~1006 limit the possible
orientations of the ISMF in the plane of the sky. Close
to the shock, the azimuthal distribution of the radio surface
brightness $S\rs{r}$ is mostly determined by $S\rs{r}(\varphi)\propto
\varsigma(\varphi)\ \sigma\rs{B}(\varphi)^{(s+1)/2}$ \citep{pet-SN1006mf}
where $\varsigma$ is the injection efficiency (a fraction of accelerated
electrons), $\sigma\rs{B}$ the compression factor for MF (unity for
parallel and 4 for perpendicular shock), $\varphi$ the azimuthal angle. If
the injection is isotropic, $\varsigma(\Theta\rs{o})=\mathrm{const}$
where $\Theta\rs{o}$ is the obliquity angle), then the bright
radio limbs correspond to projection of the equatorial belt with NW-SE
orientation of the ISMF (BarMF or barrel-like model). If, on
the other hand, the injection prefers quasi-parallel shocks, the bright
limbs of SN~1006 are two polar caps and the ISMF should be oriented
in the NE-SW direction (CapMF model).

The ISMF creates an aspect angle $\phi\rs{o}$ with the line
of sight. Recently, \cite{pet-SN1006mf} have shown that
the comparison of the experimental azimuthal profiles of the radio
brightness with those derived from theoretically synthesized radio
images can be a powerful tool to determinethe model of
electron injection, the MF orientation in the plane of the
sky and the aspect angle $\phi\rs{o}$. In particular, these
authors have shown that, under the assumptions of uniform ISMF/ISM,
the injection is be isotropic, the 3D morphology of the
remnant is BarMF and the aspect angle is $\phi\rs{o}=70^\mathrm{o}$
\citep[these results were confirmed recently by detailed
MHD calculations of][]{Reynoso2010}. For the sake of generality,
we explore in the next subsection also the CapMF model with the same
value of $\phi\rs{o}$.

\subsection{Radial profiles}
\label{SN1006cp:radio-profile}

The post-shock value (denoted hereafter by the index `s') of the
spectrum normalization $K\rs{s}$ is proportional to the injection
efficiency $\varsigma$. The injection efficiency may vary with the
shock strength (velocity). We assume that $K\rs{s}\propto V^{-b}$ where $V$
is the shock velocity, and $b$ is a parameter.

In Appendix \ref{sn1006cp:apendix_syn_brightness}, we show that the
surface brightness distribution of a Sedov SNR in the radio band is
\begin{equation}
S\rs{r}=\mathrm{const}\ {\cal S}\rs{r}(\bar
\rho,\varphi;\phi\rs{o},b)\ \nu^{-(s-1)/2}K\rs{s\|}B\rs{o}^{(s+1)/2}R
\label{sn1006cp:radio-prof}
\end{equation}
where $\bar \rho=\rho/R$, and $\rho$ is the coordinate along
the radius of the remnant $R$. ${\cal S}\rs{r}$
accounts for the evolution of the electron energy spectrum and
MF inside the SNR. For fixed $\varphi$, ${\cal S}\rs{r}(\bar
\rho)$ is an universal profile which, for a given dependence
of $\varsigma(\Theta\rs{o})$, aspect angle $\phi\rs{o}$ and index $s$,
depends only on the parameter $b$\footnote{Eq.~(\ref{sn1006cp:radio-prof})
shows also that the universal azimuthal profile of the radio
brightness ${\cal S}\rs{r}(\varphi)$ depends only on the aspect angle
$\phi\rs{o}$ because $b$ is assumed to be independent of obliquity. This
property allowed us to determine $\phi\rs{o}$ from the radio map
(see Sect.~\ref{sn1006cp:radio_azimuth})}.

\begin{figure}
\centering
\includegraphics[width=8.3truecm]{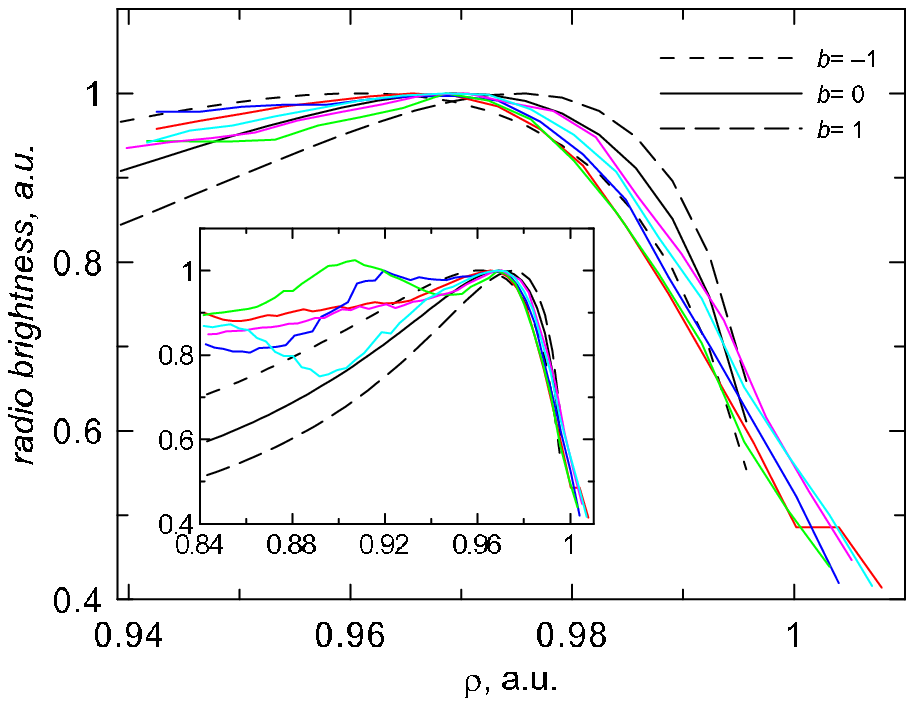}
\caption{Radial profiles of the radio brightness in NE limb of SN~1006. 
Experimental profiles (from regions 1-5, Fig.~\ref{SN1006cp:fig-a}) 
are in color. Theoretical 
profiles are in black, for $b=-1$ (dot), $b=0$ (solid), $b=1$ (dashed). 
They are calculated for $s=2$, $\phi\rs{o}=70^\mathrm{o}$, isotropic injection  
and $\varphi=70^\mathrm{o}$ 
(observational profiles are taken for $\varphi=65^\mathrm{o}-75^\mathrm{o}$).
Internal plot is a zoom-out of the main figure.} 
\label{SN1006cp:radioprofiles}
\end{figure}

We use the experimental radio image of SN~1006 presented in
\cite{pet-SN1006mf} to determine the parameter $b$. We extract
the radial profiles of radio brightness from the regions shown
in Fig.~\ref{SN1006cp:fig-a}. The profiles are reported in
Fig.~\ref{SN1006cp:radioprofiles} together with the theoretical profiles
${\cal S}\rs{r}(\bar \rho;b)$ calculated numerically for three values
of $b=-1,0,1$. Close to the shock front, the experimental profiles seems
to be between the theoretical ones calculated for $b=-1$ and $b=0$.

On the other hand, Fig.~\ref{SN1006cp:radioprofiles} shows also
that the theoretical profiles of radio brightness calculated from a
Sedov model of SNR expanding through a uniform ISM/ISMF do not fit the
experimental data to larger extent, namely for $\rho<0.94R$ (see
the inset in Fig.~\ref{SN1006cp:radioprofiles}). This result
may be explained if either the ISMF or the ISM in the neighbourhood of
SN~1006 is not uniform. In fact, the radio brightness is higher
where either the ISMF strength $B\rs{o}$ or the ISM density $n\rs{o}$
is larger: $S\rs{r}\propto \varsigma\sigma\rs{B}^{(s+1)/2}\propto
n\rs{o}B\rs{o}^{(s+1)/2}$. As a consequence, a gradient of ISMF/ISM
can cause various asymetries in the brightness distribution of the
remnant \citep{2007A&A...470..927O}; e.g. if the gradient has
a component along the line of sight, it may increase the brightness
inside the projection depending on its orientation and strength
\citep{Pet01TXC}. The radio profiles from the SW limb support
such a scenario: they monotonically increase from the shock to
$\simeq 0.85R$ (Fig.~\ref{SN1006cp:fig-a}) while the maximum of the radio
brightness in Sedov SNR should be located around $\simeq0.97R$. We
investigate the effects of a nonuniform ISMF on the remnant morphology
in a companion paper (Bocchino et al. 2010, in preparation), where
we consider an MHD model of SNR expanding through a nonuniform ISMF
and compare the synthetic images in the radio band with observations
of SN~1006. In the present study, we adopt $b=0$. 

\section{Constraints from obliquity dependence of the maximum energy}
\label{EmaxTheta0}

In this section, we aim at deriving some constraints on the modeling
of SN~1006 from the obliquity dependence of the maximum energy deduced
from the observations. \citet{SN1006Marco} considered a set of 30
regions covering the entire rim of the shell of SN~1006. The spectral
fitting of the X-ray emission extracted from these regions
allowed to derive the azimuthal variation of $\nu\rs{break}$,
a parameter in the {\sl srcut} model of {\sc XSPEC} which
is related to the maximum energy of electrons as
\begin{equation}
E\rs{max}=c_1^{-1/2}\nu\rs{break}^{1/2}B\rs{s}^{-1/2}
\label{SN1006eq1}
\end{equation}
where $B\rs{s}$ is the strength of the post-shock MF,
$c_1=6.26\E{18}\un{cgs}$. We use the above relation together with
the experimental data on $\nu\rs{break}$ to determine the azimuthal
variation of the electron maximum energy $E\rs{max}(\varphi)$.

\begin{figure}
\centering
\includegraphics[width=8.3truecm]{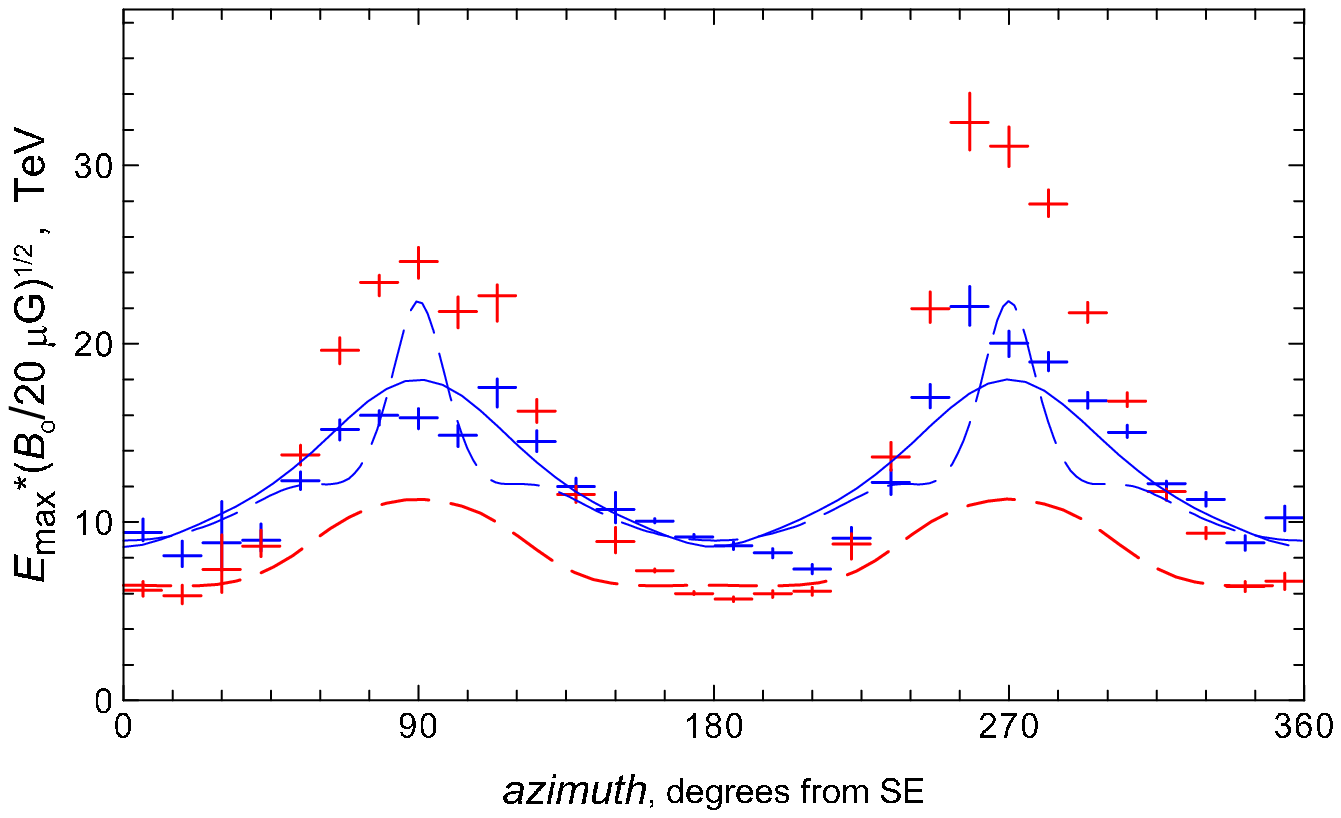} 
\caption{Variation of $E\rs{max}$ (shown with vertical one-sigma errors)
over the forward shock in SN~1006, obtained from experimental data on
$\nu\rs{break}$ \citep{SN1006Marco} and Eq.~(\ref{SN1006eq1}), for two
models of ISMF: BarMF (blue crosses), CapMF (red crosses). Aspect angle
$\phi\rs{o}=70^\mathrm{o}$.
Dashed red line: the loss-limited model with $\eta=1$, CapMF. 
Solid blue line: the time-limited model with $\eta=1.5$, BarMF. 
Dashed blue line: the loss-limited model with $\eta=7.6$, BarMF. 
              } 
\label{SN1006:Emax}
\end{figure}

The dependence of $E\rs{max}$ on the
obliquity angle $\Theta\rs{o}$ can be represented as
$E\rs{max}(\Theta\rs{o})=E\rs{max\|}f\rs{E}(\Theta\rs{o})$ where
$f\rs{E}(\Theta\rs{o})$ is a smooth function of $\Theta\rs{o}$. The
azimuthal profiles of $E\rs{max}$ determined with Eq.~(\ref{SN1006eq1})
for two different configurations of the ISMF (BarMF and
CapMF) is shown on Fig.~\ref{SN1006:Emax}.

\begin{figure}
\centering
\includegraphics[width=8.3truecm]{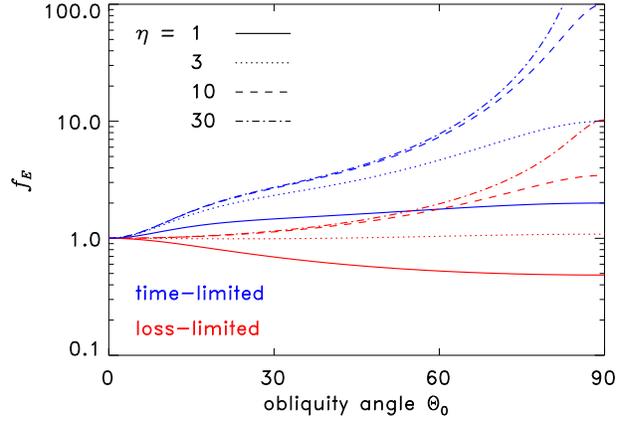}
\caption{Obliquity dependence of the function
         $f\rs{E}(\Theta\rs{o})$, determining the azimuthal dependence of
         $E\rs{max}$. The blue solid line and red solid line
         correspond to the time-limited and loss-limited model
         of \citet{Reyn-98}. Dotted and dashed lines
         correspond to the same models but using different values for the
         gyrofactor $\eta$.}
\label{obl_dep}
\end{figure}

What cause the limitation of $E\rs{max}$ in SN~1006? In the
framework of the classical theory of acceleration, \citet{Reyn-98}
(to which the reader is referred to for more details) developed
three different theoretical models for the surface variation of
$E\rs{max}$ (and, therefore, for its obliquity dependence). Namely,
the maximum energy of accelerated electrons may be determined:
1) by the electron radiative losses\footnote{Unless otherwise stated,
we consider radiative losses to be due only to the synchrotron
emission. IC losses of relativistic electrons on the cosmic microwave
background radiation (CMBR) are inefficient to produce prominent
changes in our results.} (in the following loss-limited model), 2)
by the limited time of acceleration (time-limited model), and 3) by
escaping of particles from the region of acceleration (escape-limited
model). The third model results in constant $E\rs{max}$ that contradicts
the obliquity dependence of $E\rs{max}$ derived from observations
(\citealt{SN1006Marco}; see Fig.~\ref{SN1006:Emax}). Therefore, we
do not consider it in the rest of the paper. In the other two models,
$f\rs{E}(\Theta\rs{o})$ depends basically on the MF compression ratio
$\sigma\rs{B}$, and on the level of turbulence which is reflected by
the ``gyrofactor'' $\eta\geq 1$, i.e. the ratio between the mean free
path of particle along the magnetic field and its Larmour radius (see
\citealt{Reyn-98}). Figure~\ref{obl_dep} shows the obliquity
dependence of $E\rs{max}$ in the time-limited and loss-limited models
for different $\eta$.

The obliquity angle is minimum in SN~1006 at azimuth $0^\mathrm{o}$
for BarMF and at azimuth $90^\mathrm{o}$ for CapMF. Therefore,
$E\rs{max}$ is expected to increase or decrease with obliquity
for BarMF and CapMF respectively (Fig.~\ref{SN1006:Emax}). In the
loss-limited model of $E\rs{max}$, the function $f\rs{E}(\Theta\rs{o})$
increases with increasing obliquity for $\eta\geq 3$ (red
lines in Fig.~\ref{obl_dep}). In the time-limited model,
the function $f\rs{E}(\Theta\rs{o})$ increases with obliquity
for any $\eta$ (blue lines in Fig.~\ref{obl_dep}). In
contrast, Fig.~\ref{SN1006:Emax} shows decrease of $E\rs{max}$ from
$\varphi=90^\mathrm{o}$ to $180^\mathrm{o}$ for CapMF (red crosses). Thus,
the time-limited model and loss-limited model with $\eta\geq 3$
are not applicable if one considers a polar-caps morphology of
SN~1006. In the loss-limited case, the fastest decrease with obliquity is
for $\eta=1$ but it does not fit the experimental profile of $E\rs{max}$
for model CapMF (red dashed line on Fig.~\ref{SN1006:Emax}). To the end,
the NE-SW orientation of ISMF (CapMF, polar caps) is not able to explain
observed azimuthal variation of $\nu\rs{break}$, under assumptions
of uniform ISMF/ISM and classic MHD/acceleration. We tried also other
aspect angles, $\phi\rs{o}>50^\mathrm{o}$, either with or without
the inclusion of IC radiative losses. However, the conclusion remains
unchanged.

In the BarMF case (blue crosses), the function
$f\rs{E}(\Theta\rs{o})$ for SN~1006 may be determined
by fitting the experimental data with a model of
\citet{Reyn-98}. The best-fit in the time-limited model is reached
for $\eta=1.5\pm0.02$ ($\chi^2/\mathrm{dof}=12.7$, solid blue line on
Fig.~\ref{SN1006:Emax}). The best-fit for the loss-limited model is
for $\eta=7.6\pm0.11$ ($\chi^2/\mathrm{dof}=25.8$, dashed blue line)
but the shape of the fit does not follow well the observed one.

Thus, the azimuthal variation of $\nu\rs{break}$ may be explained
in the framework of the classic MHD/acceleration theories. It
limits ISMF orientation to only BarMF configuration, in agreement
with the same conclusion obtained from azimuthal fits of the radio
surface brightness \citep{pet-SN1006mf}. The time-limited
model of \citet{Reyn-98} with $\eta=1.5$ is the most appropriate for
$E\rs{max}(\Theta\rs{o})$; we use it in the present paper. In this model,
the maximum energy of accelerated electrons varies with time very slowly
\citep{Reyn-98}. We assume therefore that $E\rs{max}$ is independent
on the shock velocity. Similar conclusions are obtained by
\citet{Katsuda2010}: the correlation they found between the X-ray flux and
the cut-off frequency is against the loss-limited model for $E\rs{max}$;
absence of time variation of the synchrotron flux supports assumption
about constant (in time) maximum energy.

The solid blue line in Fig.~\ref{SN1006:Emax} shows that
$E\rs{\max}=8.5(B\rs{o}/20\un{\mu G})^{-1/2}\un{TeV} $ at azimuth
$\varphi=0$. Since the aspect angle $\phi\rs{o}=70^\mathrm{o}$,
this value of $E\rs{\max}$ corresponds therefore to the obliquity
$\Theta\rs{o}=20^\mathrm{o}$. It is smaller at the parallel shock, namely
$E\rs{\max\|}=0.644 E\rs{\max}$ for $\eta=1.5$ \citep{Reyn-98}. Therefore,
\begin{equation}
 E\rs{\max\|}=5.4\left(\frac{B\rs{o}}{20\un{\mu G}}\right)^{-1/2}\un{TeV}.
 \label{sn1006cp:Emaxpar}
\end{equation}
The same time-limited model predicts $E\rs{\max\bot}=3.25E\rs{\max\|}$.

\section{Constraints from total radio, X-ray and TeV gamma-ray spectrum}
\label{SN1006:vol-sp}

In the calculations of the synchrotron spectrum, the self-similarity
of Sedov solutions allows us to represent the complex picture of the
synchrotron emission from the whole SNR (which includes the complicate
description of the downstream evolution of the fluid elements, the
magnetic field and the spectrum of relativistic electrons in the SNR
interior as well as the full single-electron emissivity convolved at each
point with the electron spectrum) by a single universal constant $\zeta$
and a modification factor $\eta\rs{syn}$. The former is a (reduced)
integral of the radio emissivity over the SNR volume; the latter reflects
the deviation of the X-ray spectrum from the power-law (for more details
see Appendix \ref{sn1006cp:app1}). $\eta\rs{syn}$ is defined as a
ratio of the integral (i.e. from the whole SNR) synchrotron flux at a
given frequency (e.g. at the X-rays) to the power-law extrapolation of
the radio flux to this frequency; obviously, $\eta\rs{syn}\leq 1$.

In a broad band (from radio to X-rays), the synchrotron spectrum of
the volume-integrated emission from the whole SNR may be represented by
(Appendix \ref{sn1006cp:app1})
\begin{equation}
 F\rs{syn}({\nu})=C\rs{r} \zeta {\nu}^{-(s-1)/2}
                     \eta\rs{syn}(\tilde \varepsilon;\epsilon\rs{f\|})
                     B\rs{o}^{(s+1)/2}K\rs{s\|}R^3d^{-2}
 \label{SN1006cp:Fic4syn}                     
\end{equation}
where $C\rs{r}$ is a constant, and $d$ the distance to
SNR. The constant $\zeta$ is different for different models (Appendix
\ref{sn1006cp:app1}); it is $\zeta=2.68$ for $s=2.0$ or $\zeta=2.77$
for $s=2.1$ in our reference model of SN~1006 (namely BarMF,
with isotropic injection and $b=0$).

The reduced photon energy is defined as $\tilde \varepsilon=\tilde
\nu=\nu/\nu\rs{c}(E\rs{max\|},B\rs{o})$, $\nu\rs{c}(E,B)=c_1
\left\langle\sin\phi\right\rangle E^2B$ is the synchrotron characteristic
frequency:
\begin{equation}
 \tilde \varepsilon=
                9.5 
                \ \varepsilon\rs{keV}
                \left(\frac{E\rs{max\|}}{10\un{TeV}}\right)^{-2}
                \left(\frac{B\rs{o}}{20\un{\mu G}}\right)^{-1},
\label{sn1006:tilde_e_def}
\end{equation}
where $\varepsilon\rs{keV}$ is the photon energy in keV. With
Eq.~(\ref{sn1006cp:Emaxpar}), this becomes $\tilde \varepsilon=32.6
\varepsilon\rs{keV}$. 

The reduced fiducial energy $\epsilon\rs{f}=637\left(B\rs{s}^2 t
E\rs{max}\right)^{-1}$ is one of the key parameter for modeling the X-ray
and \g-ray emission \citep{Reyn-98}. The energy $\epsilon\rs{f}$ is a
measure of the importance of radiative losses in modification of
the high-energy end of the electron spectrum and therefore of the
X-ray and \g-ray spectra and images: radiative losses are essential for
$\epsilon\rs{f}<1$; if $\epsilon\rs{f}>1$, the adiabatic losses are
dominant even for electrons with $E\sim E\rs{max}$ \citep{Reyn-98}. With
Eq.~(\ref{sn1006cp:Emaxpar}) and the age $t=1000\un{yrs}$, the
dimensionless fiducial energy at parallel shock is
\begin{equation}
 \epsilon\rs{f\|}=
 5.8\left(\frac{B\rs{s\|}}{20\un{\mu G}}\right)^{-3/2}, 
 \label{sn1006cp:etaf}
\end{equation}
where $\epsilon\rs{f\bot}$ is
$\sigma\rs{B}^2E\rs{\max\bot}/E\rs{\max\|}=52$ times smaller because
both $B\rs{s}$ and $E\rs{\max}$ are larger at the perpendicular shock.

The modification factor $\eta\rs{syn}({\nu})$ shows how the synchrotron
spectrum $F\rs{syn}({\nu})$ deviates from the power-law dependence
${\nu}^{-(s-1)/2}$. The modification factor is defined to be
$\eta\rs{syn}=1$ for the radio band; it is effective in the X-ray
band and rather quickly approaches to unity with $\nu$ decreasing
below $\nu\rs{c}(E\rs{max},B)$.

In a similar fashion, the spectral distribution of the IC emission from
the whole SNR is (Appendix \ref{sn1006cp:app3})
\begin{equation}
 F\rs{ic}(\nu)=C\rs{T} \zeta\rs{T} \nu^{-(s-1)/2} \eta\rs{ic}(\nu,\epsilon\rs{f\|},E\rs{max\|}) 
                     K\rs{s\|}R^3d^{-2}.
 \label{SN1006cp:Fic4}
\end{equation}
where $\eta\rs{ic}$ is the modification factor and $\zeta\rs{T}$
the universal constant for IC \g-rays (exact definitions are given in
Appendix \ref{sn1006cp:app3}). For our reference model of SN~1006
(BarMF, with isotropic injection, and $b=0$), $\zeta\rs{T}=0.81$
for $s=2$ and $\zeta\rs{T}=0.79$ for $s=2.1$.

\begin{figure}
\centering
\includegraphics[width=8.3truecm]{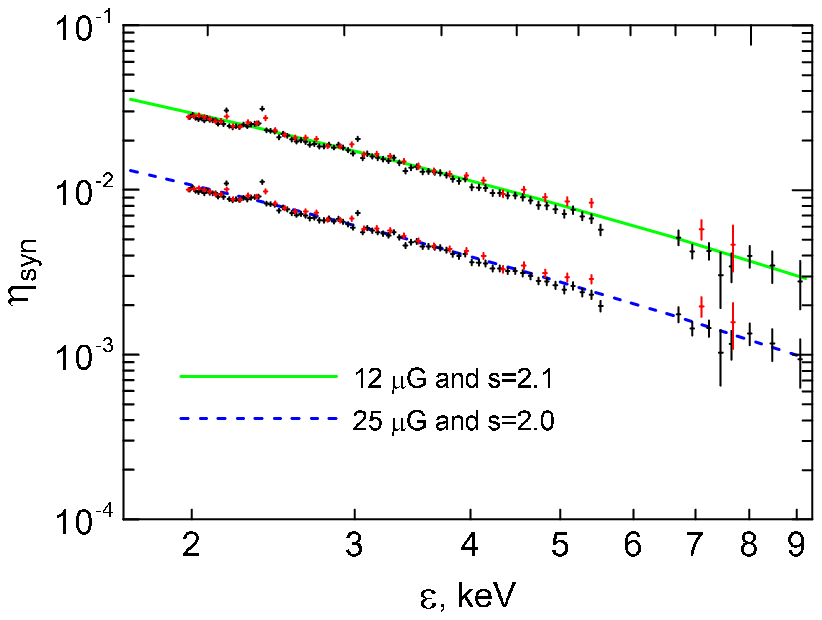}
\caption{The modification factors $\eta\rs{syn}$ in
X-rays calculated from our reference model for two sets of
parameters, namely $B\rs{o}=25\un{\mu G}$ and $s=2.0$ (dashed
blue line), and $B\rs{o}=12\un{\mu G}$ and $s=2.1$ (solid
green line). The corresponding modification factors
derived from the SUZAKU data \citep[][black and red
crosses represent XIS spectra from the front-illuminated CCDs
and back-illuminated CCD respectively]{bamba-2008-sn1006a}
are also shown for the cases $s=2.0$ (lower) and
$s=2.1$ (upper).}
\label{SN1006cp:etaXray}
\end{figure}

\subsection{Fit to the radio spectrum}
\label{SN1006cp:radio-sp}

\cite{SN1006Marco} measured the radio-to-X-ray photon index
$\alpha=(s-1)/2$ of the non-thermal component for each of the 30 regions
selected to cover the entire rim of the shell of SN~1006, finding
$\alpha\approx 0.5$. This value is almost within 1-$\sigma$ error of the
best-fit value $\alpha=0.6_{-0.09}^{+0.08}$ \citep{allen2008}, obtained
for the radio fluxes from SN~1006 at 8 different radio frequencies
\citep[most of the fluxes are from][]{Milne1971}. We consider
therefore $\alpha=0.5$ as possible choice for the spectral index of
the synchrotron spectrum. The best-fit ($\chi^2/\mathrm{dof}=1.0$)
for these radio data and fixed $\alpha=0.5$ is
\begin{equation}
 F\rs{r,obs}(\nu)=18.4
 \left({\nu}/{1\mathrm{\ \!GHz}}\right)^{-0.5}\un{Jy}.
\end{equation}
In addition, we consider also the case $\alpha=0.55$
(i.e. $s=2.1$), a value successfully used in the broad-band model of
the synchrotron and IC spectrum of SN~1006 \citep{HESS-SN1006-2010}. The
best-fit ($\chi^2/\mathrm{dof}=0.56$) for the same radio data and fixed
$\alpha=0.55$ is
\begin{equation}
 F\rs{r,obs}(\nu)=18.1
 \left({\nu}/{1\mathrm{\ \!GHz}}\right)^{-0.55}\un{Jy}.
\end{equation}

\subsection{Fit to the X-ray spectrum}
\label{sn1006cp:xray-sp}

Figure~\ref{SN1006cp:etaXray} compares the X-ray modification factor
derived from our reference model with that derived from observations
for two different sets of parameters ($B\rs{o}$, $s$). Note that the
use of the modification factor allows us to avoid uncertainties in
the distance of the remnant, its radius, and the density of emitting
electrons. The experimental modification factor $\eta\rs{syn,obs}$ is
calculated from the SUZAKU X-ray spectrum $F\rs{x,obs}$ of
the whole remnant SN~1006 \citep[Fig.~6 in][]{bamba-2008-sn1006a}
as the ratio of the observed X-ray spectrum to the extrapolation
of the radio spectrum to X-rays
\begin{equation}
 \eta\rs{syn,obs}(\nu)={F\rs{x,obs}(\nu)\over F\rs{r,obs}(\nu)}. 
\label{sn1006cp:etaxobs}
\end{equation}
This definition together with Eqs.~(\ref{sn1006:tilde_e_def})
and (\ref{sn1006cp:Emaxpar}) make the modification factor
$\eta\rs{syn,obs}(\tilde\nu)$ essentially independent on the
MF. Theoretical $\eta\rs{syn}(\tilde\varepsilon;\epsilon\rs{f\|})$,
for fixed values of $s$, $b$, $f\rs{K}(\Theta\rs{o})$ and
$f\rs{E}(\Theta\rs{o})$ is a function of the reduced fiducial energy
$\epsilon\rs{f\|}$ only. This parameter reflects the efficiency of the
radiative losses on the evolution of electrons with energies around
$E\rs{max}$ and, therefore, on the shape of the synchrotron
X-ray spectrum. In SN~1006, it is related to $B\rs{o}$ through
Eq.~(\ref{sn1006cp:etaf}).

The strength of the ambient MF $B\rs{o}=25\un{\mu G}$ together
with $s=2.0$ provide agreement between the X-ray modification factors
derived from our reference model and from the
observations (Fig.~\ref{SN1006cp:etaXray} blue line). A smaller
value of the MF strength, $B\rs{o}=12\un{\mu G}$, fits the
SUZAKU spectrum if $s=2.1$ (Fig.~\ref{SN1006cp:etaXray} green line). 

The value $B\rs{o}=25\un{\mu G}$ is close to that found in the extreme
NLA model \citep{SN1006Ber-Ksen-Volk-09}. However, NLA model assumes that
$B\rs{o}$ is compressed by the shock to the level $B\approx 150\un{\mu
G}$ and such high strength is the same everywhere in the SNR volume. In
contrast, our model allows large values of MF strength only close to
the perpendicular shock where the MF is highly compressed; as a result,
the average MF strength in the classic model of SN~1006 is smaller than
that in the extreme NLA case.

\begin{figure}
\centering
\includegraphics[width=8.3truecm]{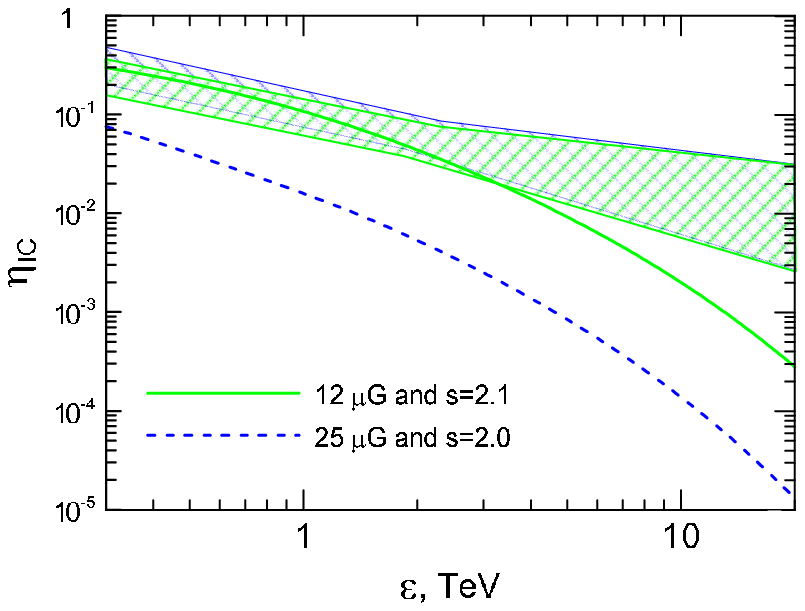}
\caption{The modification factor $\eta\rs{ic}$ in \g-rays
derived from our reference model (lines) and from the data of
\citet[][shaded regions]{HESS-SN1006-2010} for the sets of parameters
($B\rs{o}=12\un{\mu G}$, $s=2.1$) and ($B\rs{o}=25\un{\mu
G}$, $s=2.0$).}
\label{SN1006cp:etaGray}
\end{figure}

\subsection{Fit to the TeV \g-ray spectrum}
\label{sn1006cp:gamma-sp}

Fig.~\ref{SN1006cp:etaGray} compares the modification factors
$\eta\rs{ic}$ derived from our reference case and from \g-ray
observations. The experimental modification factor $\eta\rs{\gamma,obs}$
is calculated from the TeV \g-ray spectrum $F\rs{\gamma,obs}$
of SN~1006 \citep{{HESS-SN1006-2010}}.  It is evaluated as a ratio
$\eta\rs{\gamma,obs}=F\rs{\gamma,obs}/F\rs{T}$ of the observed
\g-ray spectrum to the extrapolation of the Thomson IC spectrum
to TeV \g-rays. The latter is found {from the radio spectrum} as
$F\rs{T}=\left(F\rs{T}/F\rs{r}\right)\rs{theor}F\rs{r,obs}$ where
the ratio $\left(F\rs{T}/F\rs{r}\right)\rs{theor}$ is calculated
with Eqs.~(\ref{SN1006cp:Fic4syn}) and (\ref{SN1006cp:Fic4}) for
$\eta\rs{syn}=\eta\rs{ic}=1$. Thus,
\begin{equation} 
 \eta\rs{\gamma,obs}=\frac{F\rs{\gamma,obs}}{F\rs{r,obs}}\frac{C\rs{r}\zeta}{C\rs{T}\zeta\rs{T}}
 B\rs{o}^{(s+1)/2}.
\label{sn1006cp:etagobs}
\end{equation}
The transformation of the observed TeV spectrum $F\rs{\gamma,obs}$
to the modification factor $\eta\rs{\gamma,obs}$ depends directly on
the magnetic field strength. Note that this is not a new way to estimate
the MF strength but just a different representation of the method
used by \citet{Volk-Ber-2008-MFest}.

The TeV \g-ray spectrum is almost restored by the pure
IC emission in the model with $B\rs{o}=12\un{\mu G}$ and $s=2.1$
(Fig.~\ref{SN1006cp:etaGray} green line). On the contrary, the model
with $B\rs{o}=25\un{\mu G}$ and $s=2.0$ (which is supported by the X-ray
spectrum as well; see Fig.~\ref{SN1006cp:etaXray}) does not agree with
the TeV spectrum (Fig.~\ref{SN1006cp:etaGray} blue line) if only
the leptonic \g-ray emission is considered.  Larger values of
the MF strength result in the requirement of an additional component
to fit the TeV spectrum, as it is the case in the
NLA model of \citet{SN1006Ber-Ksen-Volk-09} or in the mixed or
hadronic models of \citet{HESS-SN1006-2010}.

Since further in the present paper we consider the pure leptonic model for the
TeV \g-ray emission, we assume $B\rs{o}=12\un{\mu G}$.  It is
worth to note the difference between the spectral index $s\rs{tot}=2.1$
derived in this section for SN~1006 as a whole and those derived from
the regions covering the remnant edge, and resulting in $s\rs{loc}\approx
2.0$ \citep{SN1006Marco}. 

In Sect.~\ref{sn1006cp:xray_section}, we shall analyse the azimuthal and radial
profiles of the surface brightness extracted from regions located
quite close to the shock and use $s=2.0$, as suggested by X-ray
observations \citep{SN1006Marco}. We checked the role of other $s$, and our calculations 
(not reported here) 
show that, for $s=2.1$, the profiles of brightness are almost the same as those reported here. 

However, our calculations (not reported here) show that, for $s=2.1$, the profiles of brightness are almost the same as those reported here.

\section{Constraints from X-ray and gamma-ray maps}
\label{sn1006cp:xray_section}

\subsection{X-ray azimuthal profiles}
\label{SN1006cp:xray-az}

In Appendix \ref{sn1006cp:apendix_syn_brightness}, we demonstrate that the 
distribution of the surface brightness of a Sedov SNR due to synchrotron
emission can be represented as
\begin{equation} 
 S\rs{x}=\mathrm{const}\ {\cal S}\rs{x}(\tilde \nu,\bar \rho,\varphi;\phi\rs{o},b,\epsilon\rs{f\|})\ E\rs{max}^{1-s}K\rs{s\|}B\rs{o} R .
 \label{sn1006cp:xprof}
\end{equation}

The universal {\it shape} ${\cal S}\rs{x}$ of the radial (for fixed
$\varphi$) and azimuthal (for fixed $\bar \rho$) profiles is determined
just by one parameter, $\epsilon\rs{f\|}$, if $s$, $\phi\rs{o}$,
$b$ as well as the obliquity dependence of the injection efficiency
and the model for the maximum energy of electrons are fixed. In case
$\epsilon\rs{f\|}\gg 1$ and/or $\tilde\nu\ll 1$, the role of the radiative
losses on the downstream electron distribution is negligible and the
profiles of the brightness is then independent on the fiducial energy:
\begin{equation} 
 S\rs{x}=\mathrm{const}\ {\cal S}\rs{r}(\bar \rho,\varphi;\phi\rs{o},b)\ \tilde \nu^{-(s-1)/2}E\rs{max}^{1-s}K\rs{s\|}B\rs{o} R, 
\end{equation}
that is the same as Eq.~(\ref{sn1006cp:radio-prof}) for the radio
brightness. On the other hand, our calculations show that the role of
the evolution of injection efficiency in time (which is represented
by $b$) is less important for X-rays than for the radio because the
radiative losses (represented by $\epsilon\rs{f\|}$) are dominant in
determining the downstream distribution of X-ray emitting electrons.

\begin{figure}
\centering
\includegraphics[width=8.3truecm]{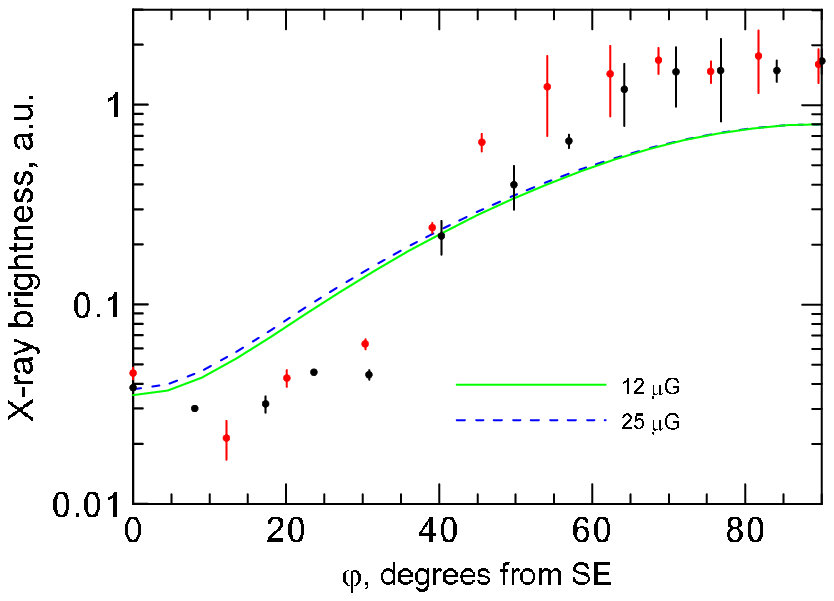}
\caption{Azimuthal profile of the X-ray brightness at fixed $\bar
\rho$ which corresponds to the maximum of the radial distribution of
brightness at $\varphi=\pi/2$. The calculations are done
for $\varepsilon=1.2\un{keV}$, $s=2$ and two values of $B\rs{o}$,
namely $12\un{\mu G}$ (solid green line) and $25\un{\mu
G}$ (dashed blue line).  The experimental data are
taken from the hard X-ray image of SN~1006 \citep{SN1006Marco}. They are
derived averaging X-ray brightness along radii within annuli centered on
the remnant (from $13.8'$ to $14.8'$ off the center for SE-NE and from
$14.4'$ to $15.2'$ for SE-SW profile). The profile from SE to SW
is in red, that from SE to NE is in black.}
\label{sn1006:fig-azXray}
\end{figure}

In our reference model of SN~1006, even $\epsilon\rs{f\|}$
is not a free parameter, because it is determined
through Eq.~(\ref{sn1006cp:etaf}) by the strength of the
MF. Fig.~\ref{sn1006:fig-azXray} compares theoretical and experimental
results. Synthesized azimuthal profiles of the X-ray brightness agree
with the observations though the fit is not ideal. Simulations reveal
that the strength of the MF is not important for the azimuthal
variation of the X-ray brightness.

Another possibility to change the azimuthal variation of the
synchrotron X-ray surface brightness in a model is to consider
a broader end\footnote{The broadening of the spectrum is
observed in SN~1006 \citep{Reyn-96,Ell-Ber-Baring-2000}; the
observed spectrum is fitted with $\alpha'=0.5-0.6$ \citep[][and
references therein]{Ell-Ber-Baring-2000}. The reason of the
broadening should be related to the property of the
acceleration process \citep{Pet06broadening} and not to
the inhomogenity of conditions in different places inside the
remnant as suggested by \citet{Reyn-96}.} of the electron spectrum {at the shock},
e.g. $N(E)\propto\exp\left(-\left(E/E\rs{max}\right)^{\alpha'}\right)$.
{The value $\alpha'=0.5\div0.6$ \citep[and references therein]{Ell-Ber-Baring-2000} 
makes the fit even worse.} Really, the azimuthal 
distribution of the brightness is roughly proportional to
$\exp\left(-(E\rs{m}(\varphi)/E\rs{max}(\varphi))^{\alpha'}\right)$ where
$E\rs{m}$ is the energy of electrons which give the largest contribution
to emission at an observed frequency. Fig~\ref{sn1006:fig-azXray}
assumes $\alpha'=1$; smaller $\alpha'$ results in smaller contrasts
between azimuth $\varphi=\pi/2$ and $\varphi=0$ that is against of the
observations. {Larger values of the parameter, $\alpha'=1\div2$, appeared in 
the loss-limited model \citep{Zirakashv-Aha-2007,Schure-et-al-2010} 
might increase the contrast but the young age of SN~1006 makes the radiative losses 
ineffective in limitation of $E\rs{max}$.}

The differences in the synthesized and observed profiles might be due
to nonuniformity of ISMF and/or ISM: larger contrasts of ISMF
or ISM density between azimuth $0^o$ and $90^o$ induced by nonuniformity
is obviously able to increase contrasts. 

\begin{figure*}
\centering
\includegraphics[height=6truecm]{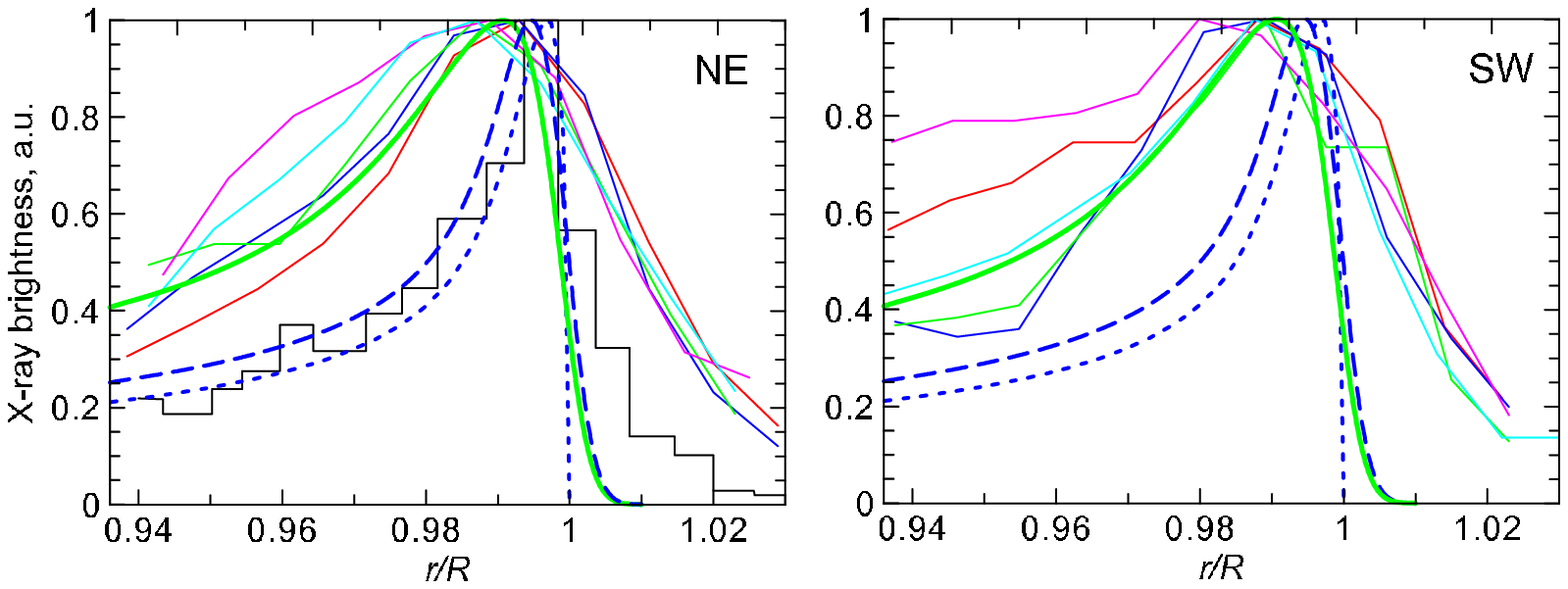}
\caption{Radial profiles of the X-ray brightness in NE (left) and SW
(right) limbs of SN~1006. Experimental XMM-Newton profiles (from
regions 1-5 and 6-10 respectively, Fig.~\ref{SN1006cp:fig-a}) for
the photon energy range $2\div4.5\un{keV}$ are in color. The sharpest
Chandra radial profile \citep[from Fig. 4A in][photon energy is $1.2\div
2\un{keV}$]{Long-et-al-2003} is shown by the histogram. Theoretical
profiles are shown by the thick blue dotted line (for $B\rs{o}=25\un{\mu
G}$, smoothed to the Chandra resolution by Gaussian with sigma $0.2''$),
and by the thick green solid line (for $12\un{\mu G}$, smoothed to the
XMM resolution by Gaussian with sigma $2.6''$). They are calculated
at 1.2 keV photons, for azimuth $\varphi=70^\mathrm{o}$ in our model of
SN~1006, $s=2$. Theoretical profile for $B\rs{o}=25\un{\mu G}$ smoothed
to the XMM-Newton resolution is shown by the long-dashed blue line.}
\label{SN1006cp:Xradial_profiles}
\end{figure*}

\subsection{X-ray radial profiles}
\label{SN1006cp:xray-profile}

The method for the MF strength estimation from the radial profile of
the X-ray brightness is described by \citet{Ber-Volk-2004-mf} \citep[see
also][]{ballet2005}. The radiative losses of electrons with energy
$E$ is $\dot{E}\propto E^2B^2$. These losses are less important for
electrons emitting in radio but they are able to modify effectively
the energy spectrum of electrons radiating X-rays. As a consequence,
the synchrotron rim in X-rays is thinner than that in radio. The
idea of the method is that the stronger the magnetic field,
the larger the radiative energy losses experienced by relativistic
electrons. This leads therefore to the rapid decrease of the spatial
distribution of electrons behind the shock and to a sharp maximum
in the radial X-ray brightness profile.  From the observational point
of view, the thinner the rim in X-rays the stronger the magnetic field
is expected to be.

Fig.~\ref{SN1006cp:Xradial_profiles} compares the theoretical profiles
${\cal S}\rs{x}(\bar \rho)$ with data from XMM and Chandra. The
simulated distribution with $B\rs{o}=12\un{\mu G}$ (green solid
line) fit the XMM data. In our model, the MF compression factor
is $\sigma\rs{B}=3.8$ at the azimuth $\varphi=70^\mathrm{o}$. The
post-shock MF is therefore $B\rs{s}\simeq 45\un{\mu G}$ in both NE and SW
limbs. This value could be considered as an upper limit for an average
MF within the limbs because some observed profiles are a bit thicker
than the theoretical one which is shown by the thick green line.
The strength $B\rs{o}=25\un{\mu G}$ (long-dashed blue line) does not
fit XMM the radial profiles of X-ray brightness.

However, the sharpest Chandra profile \citep[Fig. 4A
in][]{Long-et-al-2003} may not be explained by $B\rs{s}\simeq
45\un{\mu G}$.  Our model fits this profile if the post-shock field is
$B\rs{s}\simeq 95\un{\mu G}$ (blue dotted line). The same filament
was used by \citet{SN1006Ber-Ksen-Volk-03} to deduce $\simeq 130\un{\mu
G}$ field. Our estimate for the thinnest filament is comparable but
lower than in the NLA model. The reasons of such discrepancy are some
differences between our and their models. Namely, in our model, MF
decreases downstream of the shock while the extreme NLA model assumes
uniform MF. In addition, we accept \citep[following][]{Reyn-98}
that accelerated electrons are confined in the fluid element while
\citet{SN1006Ber-Ksen-Volk-03} include diffusion.

In general, MF estimated from the radial profile of X-ray brightness
reflects the local conditions. The quite large strength of the downstream
magnetic field, $B\simeq 130-150\un{\mu G}$, adopted in the extreme NLA
model, was assumed to be the same everywhere in the SN~1006 interior
\citep{SN1006Ber-Ksen-Volk-09}. This value is reasonable for a thinnest
NE filament \citep[Fig. 4A in][]{Long-et-al-2003} as it is apparent
from the fitting of the radial profile of X-ray brightness
\citep{SN1006Ber-Ksen-Volk-03,SN1006Ksen-Ber-Volk-05}. However,
the two close radial profiles are already thicker \citep[Fig. 4B,C
in][]{Long-et-al-2003} suggesting therefore a smaller value of $B$
even around the location of the original sharpest filament.

An effective MF inside SN~1006 (i.e. which may be used to represent SNR
as a whole) is smaller. Let us consider two possible effective field
values: the volume average
\begin{equation}
 \left\langle B\right\rangle\rs{v}=V^{-1}{\int B dV}
\end{equation}
and the radio-emissivity weighted volume average
\begin{equation}
 \left\langle B\right\rangle\rs{ev}=\int BP dV\Big/\int P dV.
\end{equation}
(The radio-emissivity weighted volume average is higher than the volume
average because the emissivity $P$ quickly decreases downstream; thus, in
calculation of the average, most of the contribution comes 
from regions close to the shock where the MF is large.) 
{It may easily be shown analytically that 
(due to self-similarity of the Sedov solution)
both $\left\langle B\right\rangle\rs{v}$ and $\left\langle B\right\rangle\rs{ev}$ 
are simply products of $B\rs{o}$ and some constants.}
Our model {predicts $B\rs{o}=12\un{\mu G}$ and} yields 
$\left\langle B\right\rangle\rs{v}=1.06B\rs{o}=12\un{\mu
G}$ and $\left\langle B\right\rangle\rs{ev}=2.7B\rs{o}=32\un{\mu G}$.
The latter is in good agreement with the strength ($\approx
30\un{\mu G}$) found in the leptonic model of \citet{HESS-SN1006-2010}
and with the estimation (again $\approx 30\un{\mu G}$) derived 
from Fig.~1 in \citet{Volk-Ber-2008-MFest} with the use of the 
HESS spectrum\footnote{{Our model deals with three-dimentional distribution 
of the magnetic field and emitting electrons while 
simpler models of \citet{HESS-SN1006-2010} and \citet{Volk-Ber-2008-MFest}  
consider uniform plasma in the uniform magnetic field.}}. 

\begin{figure}
\centering
\includegraphics[width=8.3truecm]{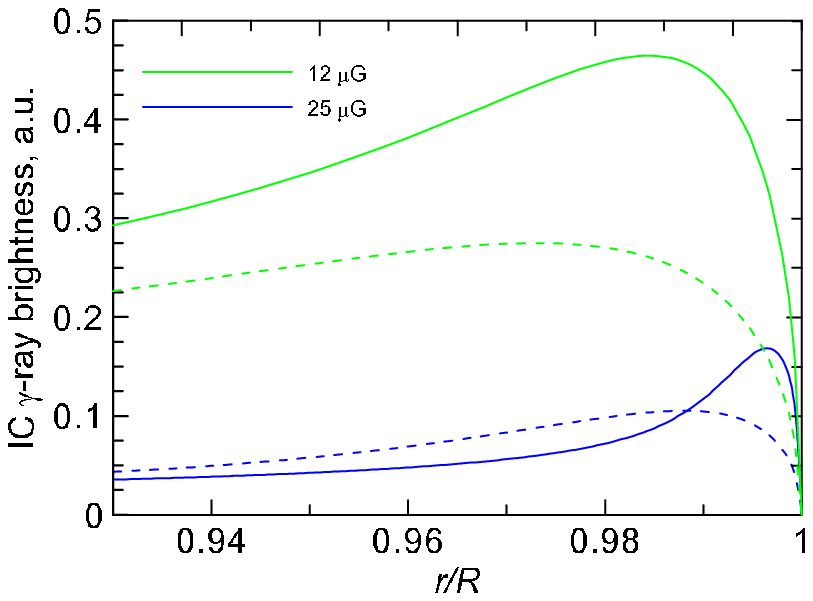}
\caption{{Radial profiles of the IC \g-ray brightness at
1 TeV in our model of SN~1006, for azimuth $\varphi=0$
(dotted lines) and $\varphi=\pi/2$ (solid lines), for
two values of $B\rs{o}$; $\phi\rs{o}=70^\mathrm{o}$,
$s=2$.} }
\label{SN1006cp:Gradial_profiles}
\end{figure}

\subsection{A note on the gamma-ray brightness}
\label{sn1006cp:gamma_section}

The distribution of the IC brightness in \g-rays with energies
$\varepsilon$ is given by (Appendix \ref{sn1006cp:apendix_IC_brightness})
\begin{equation} 
 S\rs{ic}=\mathrm{const}\ 
  {\cal S}\rs{ic}(\varepsilon,\bar \rho,\varphi;\phi\rs{o},b,\epsilon\rs{f\|},E\rs{max})\ 
  K\rs{s\|} R .
 \label{sn1006cp:gprof}
\end{equation}
This formula shows which factors affect the shape of the azimuthal and
radial profiles and which determine their amplitudes. The \g-ray
image of SN~1006 (namely, the shapes ${\cal S}$ of the azimuthal and
radial profiles in these bands) -- like the X-ray map -- depend only on
the value of $B\rs{o}$, once other parameters are fixed.

SN~1006 is rather faint in TeV \g-rays to allow, at present time,
to derive azimuthal and radial profiles with quality comparable to those
obtained in the radio and X-ray bands. However, we may check whether our
model provides the observed location of the bright \g-ray limbs.
Fig.~\ref{SN1006cp:Gradial_profiles} shows the radial profiles of the
\g-ray surface brightness in our model of SN~1006. The bright limbs
are located at the azimuth $\varphi=\pi/2$ for both strengths
of MF considered (namely, $12$ and $25\mu$G), in agreement with the
observations. Note that such property is not universal; it depends
on the parameters of the model. For example, if an aspect angle would
be $90^\mathrm{o}$ then the observed location of the TeV \g-ray limbs
may be possible only for ISMF strength larger than $\sim 100\un{\mu
G}$. We address this issue in a separate paper.

\section{Discussion and Conclusions}
\label{sn1006cp:conclusion_section}

The magnetic field strength in SN~1006 is one of the
key parameter in the model. Being related to $E\rs{max\|}$ with
Eq.~(\ref{sn1006cp:Emaxpar}) and to the parameter $\epsilon\rs{f\|}$
(which regulates efficiency of the radiative losses of relativistic
electrons) with Eq.~(\ref{sn1006cp:etaf}), it influences almost everything
in nonthermal spectra and images.

\begin{table*}
	\centering
	\caption{Summary of the observables used for parameter determination and cross-check$^\mathrm{a}$}
		\begin{tabular}{lll}
		\hline
			Observable & Parameter & Value \\
		\hline	
			radio azumuthal profile$^\mathrm{b}$ & aspect angle & $\phi\rs{o}=70^\mathrm{o}\pm4.2^\mathrm{o}$\\
			&injection type & isotropic \\
			&orientation of ISMF and {SNR morphology} & SE-NW, {barrel-like}\\
			radio radial profile & $b$ in $K\rs{s}\propto V^{-b}$ & $-1\lsim b\lsim 0$\\
			{local broad-band fits of spectra$^\mathrm{c}$}& local index $s\rs{loc}$ over shock & $s\rs{loc}=2.0$ over most of SNR rim\\
			$\nu\rs{break}$ azimuthal profile$^\mathrm{c}$ & model of $E\rs{max}$ & time-limited$^\mathrm{d}$\\
			& ratio of the mean free path to Larmour radius & $\eta=1.5$\\
			&electron maximum energy at parallel shock & $E\rs{max\|}=7.0(B\rs{o}/12\un{\mu G})^{-1/2}\un{TeV}$\\
			&electron maximum energy at perpendicular shock & $E\rs{max\bot}=3.25 E\rs{max\|}$\\
			radio and hard X-ray spectrum & MF strength {and index $s\rs{tot}$ for the whole SNR}& ($B\rs{o}=25\un{\mu G}$ and $s\rs{tot}=2.0$) or \\
			& & ($B\rs{o}=12\un{\mu G}$ and $s\rs{tot}=2.1$)\\
			radio and TeV \g-ray spectrum & \g-ray emission model, MF strength {and index $s\rs{tot}$} & IC with $B\rs{o}=12\un{\mu G}$ and $s\rs{tot}=2.1$\\
			X-ray radial profiles & post-shock MF strength in the limbs & $B\rs{s\bot}\simeq 50\un{\mu G}$\\
			X-ray azimuthal profile & MF strength & OK \\
			& ISMF orientation and aspect angle & OK \\
			& model of $E\rs{max}$ & OK \\
			\g-ray limbs location & MF strength, aspect angle & OK \\
			& \g-ray emission model & OK \\
		\hline	
		\multicolumn{3}{l}{$^\mathrm{a}$ the model assumes uniform ISMF/ISM and $\gamma=5/3$}\\
		\multicolumn{3}{l}{$^\mathrm{b}$ \citet{pet-SN1006mf,Reynoso2010}}\\
		\multicolumn{3}{l}{$^\mathrm{c}$ \citet{SN1006Marco}}\\
		\multicolumn{3}{l}{$^\mathrm{d}$ see also \citet{Katsuda2010}}\\
		\end{tabular}
	\label{sn1006cp:tab:methods}
\end{table*}

We consider a `classic' model of SN~1006, i.e. model which is based on classic MHD and acceleration theories. 
Since they are better developed compared to NLA approach, they allow us to put observational constraints on the (test-particle) kinetics and MF, to compare the azimuthal variations of the electron maximum energy and the surface brightness in radio, hard X-rays and TeV \g-rays. 
At the present time, such comparison may not be done in the frame of the NLA theory. 
We demonstrate that the `classic' model is in agreement with most of the observational data.

We try to fix free parameters of the model step-by-step, looking for observations which is mostly sensitive to some of them (Table~\ref{sn1006cp:tab:methods}). In addition to the commonly used broad-band spectrum, the properties of the nonthermal (radio, X-ray and TeV \g-ray) images of SNR {as well as spatially resolved spectral fits} are considered. 

In particular, the morphology and azimuthal profiles of the radio brightness may determine the orientation of ISMF. Namely, the radio data may be fitted by the model with uniform ISMF which is oriented perpendicular to the Galactic plane with an angle $70^\mathrm{o}$ to the line of sight \citep{pet-SN1006mf,Reynoso2010}. If so, the injection efficiency should be independent of obliquity. The radial distribution of the radio brightness depends now only on the way the injection efficiency varies with time ($K\propto V^{-b}$). The observations however may not definitively fix $b$. It is somewhere between $-1$ and $0$ but accuracy of the data allow also for a bit wider range. Spatially resolved X-ray analysis of regions around the forward shock demonstrate that distribution of $\nu\rs{break}$ may be explained by the time-limited model of $E\rs{max}$; this is in agreement with recent results of \citet{Katsuda2010}. The maximum energy of electrons at the parallel shock is found $E\rs{max\|}=7(B\rs{o}/12\un{\mu G})^{-1/2}\un{TeV}$. It is 3.25 times higher at regions where shock is perpendicular. 

We obtain expressions for the radio, X-ray and \g-ray spectra from the whole SNR in a form which clearly show which parameter of the model is responsible for the amplitude of the spectrum and which one for its shape. The modification factor of the synchrotron X-ray spectrum -- which shows the deviation of the spectrum  from the power law -- may well be explained by the classical model with ISMF strength $B\rs{o}=25\un{\mu G}$ if $s=2.0$ or with $B\rs{o}=12\un{\mu G}$ if $s=2.1$. At the same time, the TeV \g-ray modification factor prefers only the pair $B\rs{o}=12\un{\mu G}$, $s=2.1$; TeV emission is then completely due to IC process.
In case $B\rs{o}=25\un{\mu G}$, an additional component in the TeV \g-ray spectrum is needed, from pion decays, as it is in the model of \citet{SN1006Ber-Ksen-Volk-09}. 
The proton injection in such scenario should increase with obliquity in order to fit the observed azimuthal profiles of TeV \g-ray brightness 
(ISMF is parallel to the limbs). 
Could the electron and proton injections have so different dependences on obliquity in the same SNR: isotropic for electrons and quasi-perpendicular for protons?  

The extreme NLA approach \citep{SN1006Ber-Ksen-Volk-09} predicts $B\rs{o}=30\un{\mu G}$ immediately before the forward shock and $B=150\un{\mu G}$ everywhere inside the SNR. A number of the radial profiles of X-ray brightness obtained from XMM image agree with our model if an ambient MF is $B\rs{o}=12\un{\mu G}$. Around the quasi-perpendicular shock, where the profiles are extracted from, our model predicts the post-shock MF with strength $B\rs{s\bot}\simeq 50\un{\mu G}$. However, in the classic model of SN~1006, this is the value immediately post-shock; after then it rapidly decreases  downstream. Therefore, an effective (emissivity weighted average) MF within SN~1006 is estimated to be $32\un{\mu G}$ that agrees well with estimates of \cite{Volk-Ber-2008-MFest} and \citet{HESS-SN1006-2010}. MF in the sharpest Chandra profile is fitted in our model with $B\rs{s}=95\un{\mu G}$; it reflects the local conditions, only within this filament.

We found that the broad-band spectrum from the whole SN~1006 is better represented with the electron spectral index $s\rs{tot}=2.1$ while local radio-to-X-ray spectra over the SNR shock prefers $s\rs{loc}=2.0$ \citep{SN1006Marco}. 
{It is interesting, that similar difference in spectral index is found in the theoretical study of \citet{Schure-et-al-2010}: 
the spectrum near the shock is flatter than the overall spectrum. The authors attributed this difference to the time evolution of $E\rs{max}$ 
which was lower at previous times. In contrast, the time-limited model for the maximum energy \citep{Reyn-98} which fits the azimuthal variation of $\nu\rs{break}$  in SN~1006 (Sect.~\ref{EmaxTheta0}) as well as absence of the time variation of the synchrotron flux \citep{Katsuda2010}, suggest what $E\rs{max}$ varies quite slowly in this SNR, at least in the regions close to the shock. The issue of different spectral slopes has to be considered in the future. As to the purpose of the present study, }
difference between $s\rs{tot}$ and $s\rs{loc}$ is negligible for azimuthal and radial profiles of radio, X-ray and IC \g-ray brightness.

Azimuthal profiles of the X-ray and \g-ray brightness in our model behave in the same way as in the observations.

`Classic' model has also few difficulties. 
\citet{Rotetal04} developed a simple geometrical criterion to distinguish between barrel-like and polar-cap morphology in SN~1006. They have shown analytically that the ratio between the central and the rim brightness should be larger than some value in BarMF case (projected ``barrel'' has to provide enough brightness in the internal regions). In XMM map, this ratio is smaller. The luck of brightness from equatorial belt in the central part is an argument against BarMF morphology. 
Our model, which strongly prefer BarMF, does not agree with the criterion of \citet{Rotetal04}. Nevertheless, the polar-cap scenario, which is in agreement with this criterion (and is adopted by the NLA model), is unable to explain the observed azimuthal profiles of the break frequency $\nu\rs{break}$and the radio brightness, under assumptions that ISMF/ISM are uniform and the amplified/compressed MF increases with obliquity. 

Another minor point of the classic model of SN~1006 is the rather large ambient MF, $B\rs{o}=12\un{\mu G}$, which is difficult to expect without MF amplification at the high location of SN~1006 above the Galactic plane. 

Our model deals with ideal gas with the adiabatic index $\gamma=5/3$ and cannot explain the small distance between the forward shock and the contact discontinuity \citep{chr08,SN1006Marco}. Instead, if acceleration is so efficient that relativistic particles affect hydrodynamics then the adiabatic index may be smaller than ours. The small distance observed may naturally be explained by such, more compressible, plasma with the index like $\gamma=1.1$ \citep{orland-et-al-2010}. It is worth noting that in such case of efficient acceleration, there is no need for MF amplification: the only shock compression to factor $\sigma=21$ (as it is for $\gamma=1.1$) may result in quite large downstream MF even in case of $B\rs{o}$ of few $\un{\mu G}$. 

The two models, classical and extream NLA, are compared in Table~\ref{sn1006cp:tab:ModelsOfSN1006}. It is evident that none of them explaine the whole set of the SN~1006 properties. A new model of SN~1006 has to include either combination of the two extremes or inclusion of the ISMF/ISM nonuniformity. 

All results presented here are obtained under assumption that SN~1006 evolve in the uniform ISMF and uniform ISM. 
It is shown that the scenario of classic MHD/acceleration plus uniform ISMF/ISM strongly prefers the barell-like morphology of SN~1006.
However, we also see that nonuniform ISMF/ISM could be an essential element in the model of SN~1006. 
In particular, slanted lobes, the inversion of the brightness ratio between NE and SW limbs from radio to X-ray band and the higher break frequency in NE limb may only be explained by presence of gradient of ISMF and/or ISM.
We expect that the effect of the nonuniform ISMF might dominate the role of some nonlinear effects arising from efficient acceleration of cosmic rays
 by the forward shock in SN~1006. We like to address this issue in the future.

\begin{table*}	
	\centering
	\caption{Comparison of SN~1006 models}
	\begin{tabular}{l|l|l}
			\hline
			Properties of SN~1006&Classic model$^\mathrm{a}$&Extream NLA model$^\mathrm{b}$\\			
				&(with uniform ISMF) &(with uniform ISMF)\\
			\hline
			two-limbs in radio and hard X-ray image & YES  &YES? \\
				&\hspace{0.2cm}SN~1006 is barrel         &\hspace{0.2cm}SN~1006 has polar caps\\
				&\hspace{0.2cm}ISMF direction: SE-NW &\hspace{0.2cm}ISMF direction: SW-NE\\
			location of TeV \g-ray limbs &YES &YES?\\
			\citet{Rotetal04} criterion	&NO &YES\\
			radio spectrum &YES as power low &YES with concave shape\\
			hard X-ray spectrum &YES with $\left\langle B\right\rangle\rs{ev}=32\un{\mu G}$ 
								&YES with $\left\langle B\right\rangle\rs{v}\approx 150\un{\mu G}$\\
			TeV \g-ray spectrum &YES &YES \\
				&\hspace{0.2cm}IC with $\left\langle B\right\rangle\rs{ev}\approx 32\un{\mu G}$ 
								&\hspace{0.2cm}IC with $\left\langle B\right\rangle\rs{v}\approx 150\un{\mu G}$ \\
&
								&\hspace{0.2cm} and hadronic component\\
			radio radial profile &YES &NO? (uniform $B$ inside SNR)\\
			{sharpest X-ray radial profile} &YES with $B\rs{s}\approx95\un{\mu G}$
								&YES with $B\rs{s}\approx 150\un{\mu G}$\\
			radio azimuthal profile&YES &?\\
			hard X-ray azimuthal profile &YES &?\\
			TeV \g-ray azimuthal and radial profiles &YES &? \\
			$\nu\rs{break}$ azimuthal profiles &YES &NO? \\
			pre-shock MF strength $B\rs{o}=12\un{\mu G}$& NO &YES\\
				&\hspace{0.2cm}(if ISMF around SN~1006 &\hspace{0.2cm}as result of amplification\\
				&\hspace{0.2cm} is typical $\sim3\un{\mu G})$&\hspace{0.2cm}(if any)\\
			(eventual) concave shape of the radio spectrum$^\mathrm{c}$&NO&YES\\
			very close forward shock and contact discontinuity&NO&YES (with $\gamma=1.1$)\\
			radio `overbrightness' in the SNR interior &NO &NO? \\			
			slanted lobes &NO &NO\\
			ratio of radio ${\cal R}\rs{r}<1$ and X-ray brightness ${\cal R}\rs{x}>1$ &NO &NO\\
			$\nu\rs{break,NE}/\nu\rs{break,SW}>1$&NO &NO\\
			\hline
		{$^\mathrm{a}$} present paper\\
		{$^\mathrm{b}$} \citet{SN1006Ber-Ksen-Volk-09}\\
		{$^\mathrm{c}$} \citet{allen2008}
		\end{tabular}
	\label{sn1006cp:tab:ModelsOfSN1006}
\end{table*}

\section*{Acknowledgments}
Aya Bamba is greatly acknowledged for providing the data on SUZAKU X-ray spectrum of SN1006.
OP was partially supported by the program 'Kosmomikrofizyka' (NAS of Ukraine).


\appendix
\section[]{Surface brightness of Sedov SNR}

Surface brightness of a spherical SNR is an integral of volume emissivity $q$ along the line of sight
\begin{equation}
 S\rs{syn}=2\int\limits_{0}^{R}qdl=2R\int\limits_{\bar a(\bar\rho)}^{1}q \frac{\bar r\bar r_{\bar a}d\bar a}{\sqrt{\bar r^2-\bar\rho^2}},
\end{equation}
where $\rho$ is distance from the center of projection, $\bar r=r/R$, $a$ Lagrangian coordinate, $r_a=dr/da$, 
\begin{equation}
 q=\int N(E)p(E,\varepsilon)dE,
\end{equation}
where $E$ and $\varepsilon$ are the electron and photon energies, $p$ the radiation power of a single electron. In case of Sedov SNR in uniform medium the electron energy distribution downstream of the shock is \citep{Pet-Beshl-en-2008}
\begin{equation}
 N(E)=KE^{-s}{\cal E}\rs{rad}^{s-2}\exp\left(-\frac{E
 }{E\rs{max\|}{\cal E}\rs{ad}{\cal E}\rs{rad}f\rs{E}}\right),
\end{equation}
the normalization $K=K\rs{s\|}(t)f\rs{K}(\Theta\rs{o})\bar K(\bar a)$, the magnetic field $B=B\rs{s\|}(t)\sigma\rs{B}(\Theta\rs{o})\bar B(\bar a)$ and the electron maximum energy $E\rs{max}=E\rs{max\|}f\rs{E}(\Theta\rs{o})$.

\subsection[]{Synchrotron emission}
\label{sn1006cp:apendix_syn_brightness}

The synchrotron radiation power is
\begin{equation}
 p=\frac{\sqrt{3}e^3\left\langle\sin\phi\right\rangle}{m\rs{e}c^2}BF\rs{syn}\left(\frac{\nu}{\nu\rs{c}}\right),
\end{equation}
where all notations have their common meaning.
The synchrotron surface brightness of Sedov SNR is therefore
\begin{equation}
 S\rs{syn}=\frac{2\sqrt{3}e^3\left\langle\sin\phi\right\rangle}{m\rs{e}c^2}
  {\cal S}\rs{syn}(\tilde \nu,\bar \rho,\varphi;\phi\rs{o},b,\epsilon\rs{f\|})\ E\rs{max}^{1-s}K\rs{s\|}B\rs{o} R .
  \label{sn1006app:Ssyn}
\end{equation}
where ${\cal S}\rs{syn}(\tilde \nu,\bar \rho,\varphi)$ is a universal dimensionless function 
\begin{equation}
 \begin{array}{l}
 \displaystyle
 {\cal S}\rs{syn}=\int\limits_{\bar a(\bar \rho)}^{1}
 \left[
 \int\limits_{0}^{\infty}F\rs{syn}\left(\frac{\tilde\nu}{\epsilon^2\sigma\rs{B}\bar B}\right)\epsilon^{-s}
 {\cal E}\rs{rad}^{s-2}
 \right.\\ \\
 \displaystyle \qquad\qquad\times\left.
 \exp\left(-\frac{\epsilon
 }{{\cal E}\rs{ad}{\cal E}\rs{rad}f\rs{E}}\right)
 d\epsilon
 \right]
 \\ \\
 \displaystyle \qquad\qquad\times
 \sigma\rs{B}\bar B f\rs{K}\bar K
 \frac{\bar r\bar r_{\bar a}d\bar a}{\sqrt{\bar r^2-\bar\rho^2}},
 \end{array}
\end{equation}
where $\epsilon=E/E\rs{max\|}$. It depends on the dimensionless models of obliquity variations of $K$, $B$ and $E\rs{max}$ (i.e. on $f\rs{K}$, $\sigma\rs{B}$, $f\rs{E}$) but is independent of the actual values of $E\rs{max}$, $K\rs{s}$, $B\rs{o}$ and $R$.

In the limit $\epsilon\rs{f\|}\gg 1$ and/or $\tilde \nu\ll 1$, Eq.~(\ref{sn1006app:Ssyn}) transforms to 
\begin{equation}
\begin{array}{l}
 S\rs{syn}=
 \displaystyle
 \frac{2\sqrt{3}e^3\left\langle\sin\phi\right\rangle{\cal A}(s)}{m\rs{e}c^2}\ 
  {\cal S}\rs{r}(\bar \rho,\varphi;\phi\rs{o},b)\\ \\
  \qquad\times\ 
  {\tilde\nu}^{-(s-1)/2}
  E\rs{max}^{1-s}K\rs{s\|}B\rs{o} R .
  \end{array}
 \label{sn1006app:Sra}
\end{equation}
where 
\begin{equation}
 {\cal S}\rs{r}=\int\limits_{\bar a(\bar \rho)}^{1}\left(\sigma\rs{B}\bar B\right)^{(s+1)/2}f\rs{K}\bar K
 \frac{\bar r\bar r_{\bar a}d\bar a}{\sqrt{\bar r^2-\bar\rho^2}},
\end{equation}
or, in other form,
\begin{equation}
\begin{array}{l}
 S\rs{r}=
 \displaystyle
 \frac{2\sqrt{3}e^3\left\langle\sin\phi\right\rangle{\cal A}(s)}{m\rs{e}c^2}\ 
  {\cal S}\rs{r}(\bar \rho,\varphi;\phi\rs{o},b)\\ \\ 
  \qquad\times\ 
  \left(\nu/c_1\right)^{-(s-1)/2}
  K\rs{s\|}B\rs{o}^{(s+1)/2} R .
  \end{array}
 \label{sn1006app:Srb}
\end{equation}

\subsection[]{IC emission}
\label{sn1006cp:apendix_IC_brightness}

The IC radiation power is
\begin{equation}
 p=\frac{2e^4m\rs{e}^2c^2kT}{\pi \hbar^3}E^{-2}{\cal I}(E,\varepsilon),
\end{equation}
where all notations have their common meaning, ${\cal I}$ is a special integral \citep[see e.g.][]{Pet08IC}.
The IC brightness is therefore
\begin{equation}
 S\rs{ic}=\frac{4e^4m\rs{e}^2c^2kT}{\pi \hbar^3}\ 
  {\cal S}\rs{ic}(\varepsilon,\bar \rho,\varphi;\phi\rs{o},b,\epsilon\rs{f\|},E\rs{max})\ 
  K\rs{s\|} R .
  \label{sn1006app:Sic}
\end{equation}
The function ${\cal S}\rs{ic}(\bar \rho,\varphi)$ is not so universal as in case of the synchrotron emission; it depends on the absolute values of 
the photon energy and the maximum electron energy; we do not present it here.

\section[]{Nonthermal spectrum of Sedov SNR}

Flux is defined as 
\begin{equation}
 F(\nu)=\left(4\pi d^2\right)^{-1}\int P(\nu)dV
 \label{SN1006cp:fluxdef}
\end{equation}
where $V$ is the volume of SNR and $P$ the volume emissivity. 
We assume that the energy spectrum of electrons in the form 
\begin{equation}
 N(E)dE=KE^{-s}\exp(-E/E\rs{max})dE
 \label{el-sp-app}
\end{equation}
are created at the shock. The volume emissivity is
\begin{equation}
 P(\nu)=\int N(E)p(E,\nu)dE
\end{equation}
where $p$ is the spectral distribution of radiation power of `single' electron with energy $E$. 
Let us consider adiabatic SNR in uniform ISM and uniform ISMF. 

In general, the efficiency of injection may depend on the shock obliquity angle $\Theta\rs{o}$. 
If particles are injected easier at quasiparallel shocks then $K\rs{s}(\Theta\rs{o})$ is decreasing function of 
$\Theta\rs{o}$ with decrement rate dependent on the level of turbulence, shock strength etc. \citep{ell-bar-jones-95}. 
Let us consider parametric representation $K\rs{s}=K\rs{s\|}f\rs{K}(\Theta\rs{o})$ with approximation 
$
 f\rs{K}=\exp\left(-\left(\Theta\rs{o}/\Theta\rs{K}\right)^2\right)
$
where $K\rs{s\|}$ the normalization for region immediately after the parallel shock, $\Theta\rs{K}$ the parameter. $\Theta\rs{K}=\pi/6$ approximates the classical quasiparallel dependence, $\varsigma\propto \cos^2(\Theta\rs{o})$. 
In case of the isotropic injection, $\Theta\rs{K}=\infty$.

\begin{figure}
\centering
\includegraphics[width=7.6truecm]{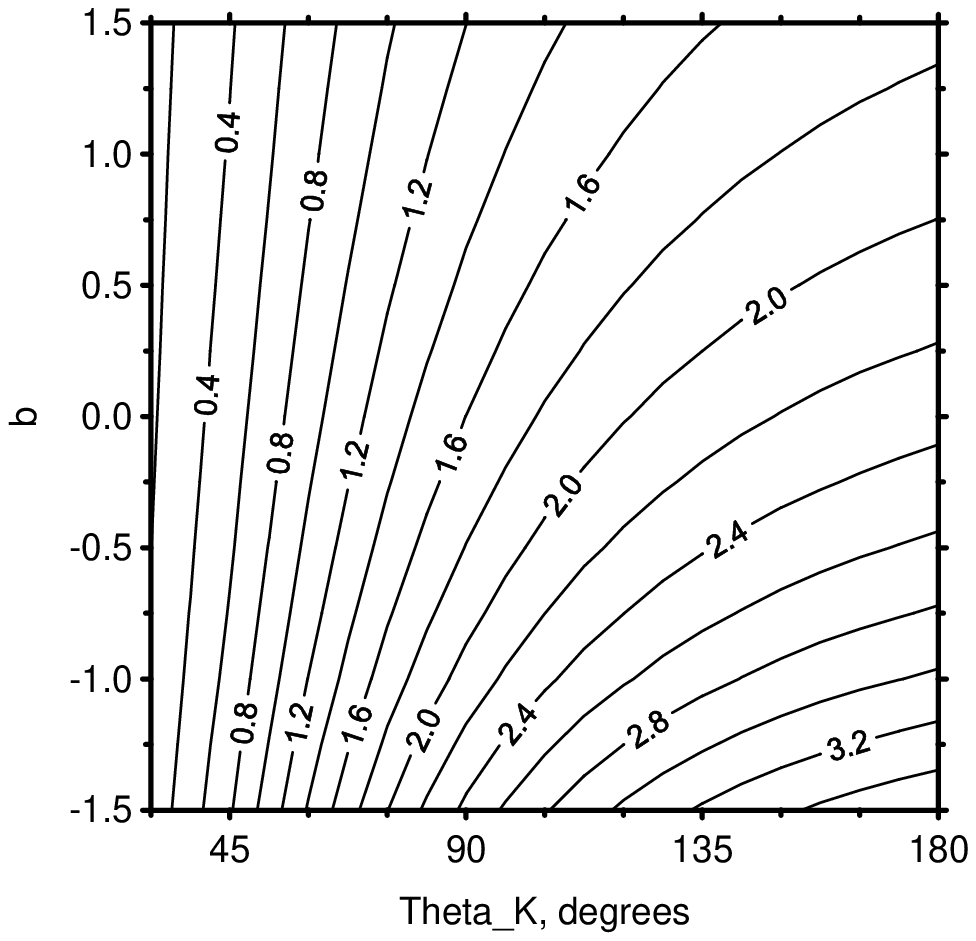} 
\caption{$\zeta$ for different values of parameters $b$ and $\Theta\rs{K}$. $s=2$
              } 
\label{SN1006cp:zeta}
\end{figure}
\begin{figure}
\centering
\includegraphics[width=8.3truecm]{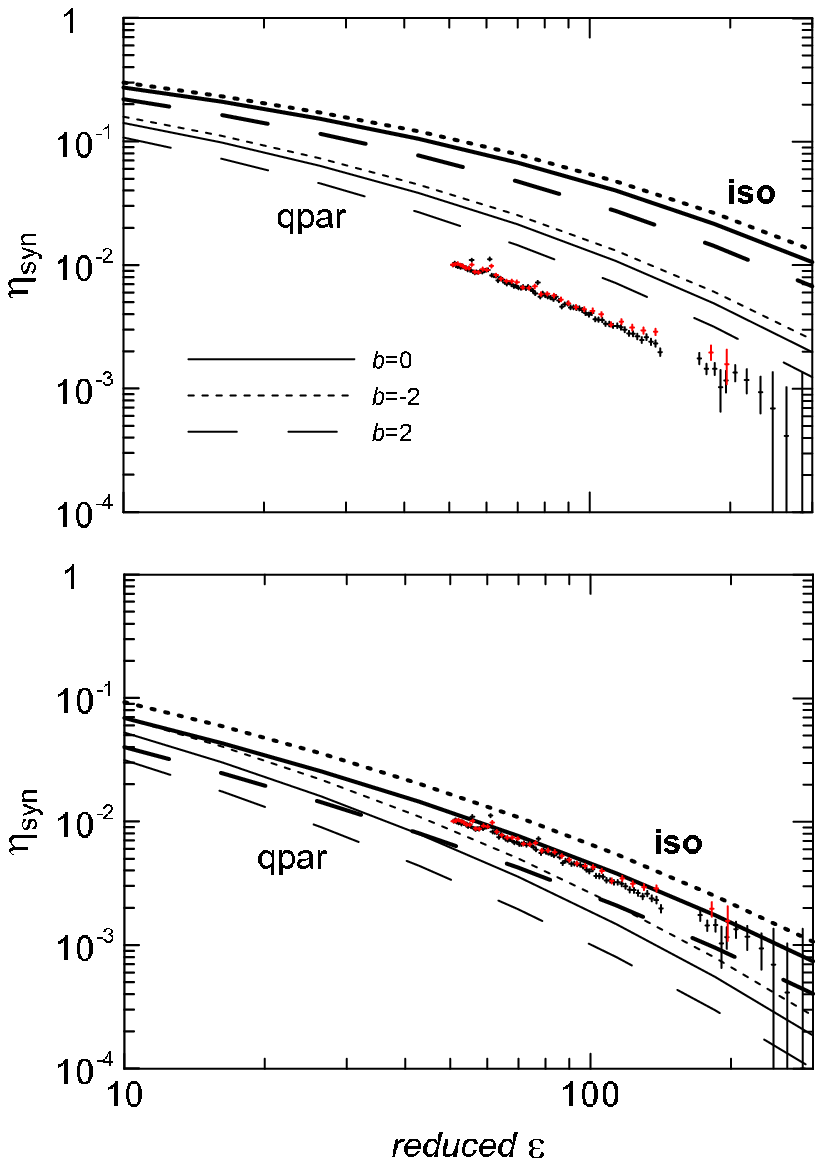} 
\caption{Modification factor $\eta\rs{syn}$. Calculations are done for $s=2$, the  
time-limited model of $E\rs{max}$ with $\eta=1.5$, 
isotropic injection (thick lines) and quasiparallel injection (thin lines), three values of $b$, 
$\epsilon\rs{f\|}=100$ (upper panel) and $\epsilon\rs{f\|}=3.2$ (lower panel). 
Experimental modification factor for SN~1006 are shown for comparison. 
It is obtained from the SUZAKU spectrum \citep[][Fig.~6]{bamba-2008-sn1006a} for photon energies $\geq 2\un{keV}$,
with the use of Eq.~(\ref{sn1006cp:etaxobs}). 
MF strength is given by Eq.~(\ref{sn1006cp:etaf}): $B\rs{o}=3\un{\mu G}$ (upper panel) and 
$B\rs{o}=30\un{\mu G}$ (lower panel).
              } 
\label{SN1006cp:etaXfig}
\end{figure}

\subsection{Synchrotron emission}
\label{sn1006cp:app1}

The radio flux (\ref{SN1006cp:fluxdef}) from Sedov SNR may be written as \citep[for details, see][]{Pet-Beshl-en-2007}
\begin{equation}
 F\rs{r}(\nu)=C \nu^{-(s-1)/2} \zeta(b,\Theta\rs{K}) K\rs{s\|}B\rs{o}^{(s+1)/2}R^3d^{-2}
 \label{SN1006cp:Fr1}
\end{equation}
where 
\begin{equation}
 C={(4\pi)^{-1}}{{\cal A}(s)c_2\mu\rs{\phi} c_1^{(s-1)/2}}, 
\end{equation}
$c_1=3e/(4\pi m\rs{e}^3c^5)$, $c_2=\sqrt{3}e^3/(m\rs{e}c^2)$, 
\begin{equation}
 {\cal A}(s)=\frac{2^{(s-1)/2}}{s+1}\Gamma\left(\frac{3s+19}{12}\right)\Gamma\left(\frac{3s-1}{12}\right),
\end{equation}
$\mu\rs{\phi}=\left\langle \sin(\varphi)^{(s+1)/2}\right\rangle$, ($C=3.493\E{-14}\un{cgs}$ in case $s=2$), 
$\varphi$ the angle between MF and the line of sight, 
\begin{equation}
 \zeta(b,\Theta\rs{K})=
  \int\limits_{0}^{2\pi}\!d\varphi\!
 \int\limits_{0}^{\pi}\!\!d\theta\sin\theta f\rs{K}\!
 \int\limits_{0}^{1}\!\!d\bar a \bar r^{2} \bar r\rs{\bar a} 
 \bar K\! \left(\sigma\rs{B}\bar{B}\right)^{(s+1)/2}\!\!,
\end{equation}
$\sigma\rs{B}(\Theta\rs{o})$ is the compression factor for MF, 
$r$ and $a$ are Eulerian and Lagrangian coordinates respectively, $r\rs{a}=dr/da$, bar represents parameter divided by its post-shock value, $(\varphi,\theta)$ spherical coordinates. 
Thanks to the self-similarity, the constant $\zeta$ `compactifies' the whole downstream evolution of fluid elements \citep{Sedov-59}, magnetic field and relativistic electrons \citep{Reyn-98}. 

In a similar fashion, the X-ray flux is \citep{Pet-Beshl-en-2008}
\begin{equation}
 F\rs{x}(\tilde \nu)=C_2 \zeta\rs{x}(\tilde \nu;b,\Theta\rs{K},\epsilon\rs{f\|})   
                     K\rs{s\|}B\rs{o}E\rs{max\|}^{1-s}R^3d^{-2}
 \label{SN1006cp:Fx1}
\end{equation}
where 
$\tilde \nu=\nu/\nu\rs{c}(E\rs{max\|},B\rs{o})$, $\nu\rs{c}(E,B)\propto E^2B$ is the synchrotron characteristic frequency, $C_2=c_2\left\langle \sin\varphi\right\rangle/(4\pi)$ a constant, $\epsilon\rs{f\|}=637\left(B\rs{s\|}^2 t E\rs{max\|}\right)^{-1}$ is the reduced fiducial energy. 
The energy $\epsilon\rs{f}$ is a measure of importance of radiative losses in modification of the 
electron spectrum 
\citep{Reyn-98}.
The function 
\begin{equation}
\begin{array}{l}
 \zeta\rs{x}(\tilde{\nu};b,\Theta\rs{K},\epsilon\rs{f\|})=
 \displaystyle
 \int\limits_{0}^{2\pi}\!d\varphi\!
 \int\limits_{0}^{\pi}\!\!d\theta\sin\theta f\rs{K}\!
 \int\limits_{0}^{1}\!\!d\bar a \bar r^{2} \bar r\rs{\bar a} 
 \bar K \sigma\rs{B}\bar B\! 
 \\ \\ \quad \times\displaystyle
 \int\limits_{0}^{\infty}\!d\epsilon \epsilon^{-s}{\cal E}\rs{rad}^{s-2}
 \exp\left(-\frac{\epsilon
 }{{\cal E}\rs{ad}{\cal E}\rs{rad}f\rs{E}}\right)
 F\rs{syn}\left(\frac{\tilde{\nu}}{\epsilon^2\sigma\rs{B}\bar B}\right),
\end{array} \!\!\!\!\!
 \label{eta-f-la2}
\end{equation}
where ${\cal E}\rs{ad}(a)$, ${\cal E}\rs{rad}(a;\epsilon\rs{f\|},\Theta\rs{o})$ represent adiabatic and radiative losses of relativistic electrons \citep{Pet-Beshl-en-2008}, $F\rs{syn}$ the function known in the theory of synchrotron radiation, $\epsilon=E/E\rs{max\|}$.

With $\tilde \nu$, the radio flux (\ref{SN1006cp:Fr1}) may be written in a form similar to (\ref{SN1006cp:Fx1}):
\begin{equation}
 F\rs{r}(\tilde \nu)=C_2 {\cal A}(s) \tilde \nu^{-(s-1)/2} \zeta(b,\Theta\rs{K})
                     K\rs{s\|}B\rs{o}E\rs{max\|}^{1-s}R^3d^{-2}.
 \label{SN1006cp:Fr2}                     
\end{equation}
Comparison of (\ref{SN1006cp:Fx1}) and (\ref{SN1006cp:Fr2}) demonstrates that, 
for $\nu$ much smaller than X-ray frequencies, $\zeta\rs{x}$ transforms to $\zeta$, as expected:
\begin{equation}
 \zeta\rs{x}(\tilde{\nu})= {\cal A}(s)\ 
 \tilde{\nu}^{-(s-1)/2}\zeta.
 \label{SN1006:eqzeta}
\end{equation}
This transition may also be shown analytically from (\ref{eta-f-la2}), in the limit $E\ll E\rs{max}$ and $E\ll \epsilon\rs{f}E\rs{max}$ \citep{Pet-Beshl-en-2008}. 

Let us introduce the modification factor for the synchrotron spectrum 
\begin{equation}
 \eta(\tilde \nu,\epsilon\rs{f\|})=\frac{\zeta\rs{x}(\tilde{\nu},\epsilon\rs{f\|})\tilde{\nu}^{(s-1)/2}}
 {{\cal A}(s)\zeta}.
 \label{modfsynch}
\end{equation}
It is defined to be $\eta\leq 1$ and ensure $\eta\rightarrow 1$ for $\nu\ll \nu\rs{c}(E\rs{max\|},B\rs{o})$, as it is given by (\ref{SN1006:eqzeta}). 
In terms of $\tilde\nu$, the modification factor is almost universal (i.e. allows for scaling with frequency). 

With the modification factor, the expression (\ref{SN1006cp:Fx1}) which describes the broad-band (radio-to-X-ray) synchrotron spectrum from Sedov SNR becomes  
\begin{equation}
 F(\nu)=C \nu^{-(s-1)/2} \zeta(b,\Theta\rs{K})
                     \eta(\tilde \nu;\epsilon\rs{f\|})
                     K\rs{s\|}B\rs{o}^{(s+1)/2}R^3d^{-2}.
\end{equation}

The values of $\zeta$ are shown on Fig.~\ref{SN1006cp:zeta}. The parameter $\zeta$ is important in normalization of synchrotron spectrum: it varies in about 8 times over the parameter space. If injection is considerably larger at parallel shocks ($\Theta\rs{K}\leq \pi/3$), the value of $b$ is almost unimportant for amplitude of the synchrotron spectrum, but rather small changes in $\Theta\rs{K}$ may cause differences in $\zeta$ in few times. In contrast, if injection tends to be isotropic ($\Theta\rs{K}\geq 2\pi/3$), $b$ plays the dominant role. 

In order to explore the parameter space, we made several runs to calculate the modification factors for different sets of parameters. Results are shown on 
Fig.~\ref{SN1006cp:etaXfig} where we also plot the experimental data in order to demonstrate relevance of the parameters for SN~1006. The modification factor  depends on $\epsilon\rs{f\|}$, $b$, $\Theta\rs{K}$ and $s$ as well as on the function $f\rs{E}(\Theta\rs{o})$. 

\subsection[]{Inverse-Compton emission}
\label{sn1006cp:app3}

The inverse-Compton flux (\ref{SN1006cp:fluxdef}) from electrons in a black-body photon field with temperature $T\rs{CMB}$, at photon energies far below TeV (i.e. when the Thomson regime and power-law electron distribution are assumed, see \citet{Pet08IC} for details), is  
\begin{equation}
 F\rs{T}(\varepsilon)=C\rs{T} \varepsilon^{-(s-1)/2} \zeta\rs{T}(b,\Theta\rs{K})   
                     K\rs{s\|}R^3d^{-2}
 \label{SN1006cp:Fic1}
\end{equation}
where $\varepsilon$ is the photon energy, 
\begin{equation}
 \zeta\rs{T}(b,\Theta\rs{K})=
  \int\limits_{0}^{2\pi}\!d\varphi\!
 \int\limits_{0}^{\pi}\!\!d\theta\sin\theta f\rs{K}\!
 \int\limits_{0}^{1}\!\!d\bar a \bar r^{2} \bar r\rs{\bar a} 
 \bar K,
\end{equation}
reflects the evolution of relativistic electrons downstream and 
\begin{equation}
 C\rs{T}=
  \frac{2^{s-1}\pi^2\sigma\rs{T}m\rs{e}{\cal A}\rs{T}(s)^{(s+1)/2}\left(kT\rs{CMB}\right)^{(s+5)/2}}
  {(s+1)h^3(m\rs{e}c^{2})^s} 
\end{equation}
where $\sigma\rs{T}$ is the Thomson cross-section, 
{\small
\begin{equation}
 {\cal A}\rs{T}(s)=
 \left[
 \frac{12}{\pi^2}\frac{(s^2+4s+11)}{(s+5)(s+3)^2}
 \int\limits_{0}^{\infty}\frac{z^{(s+3)/2} dz}{\exp(z)-1}
 \right]^{2/(s+1)}. 
\end{equation}
}

\begin{figure}
\centering
\includegraphics[width=7.6truecm]{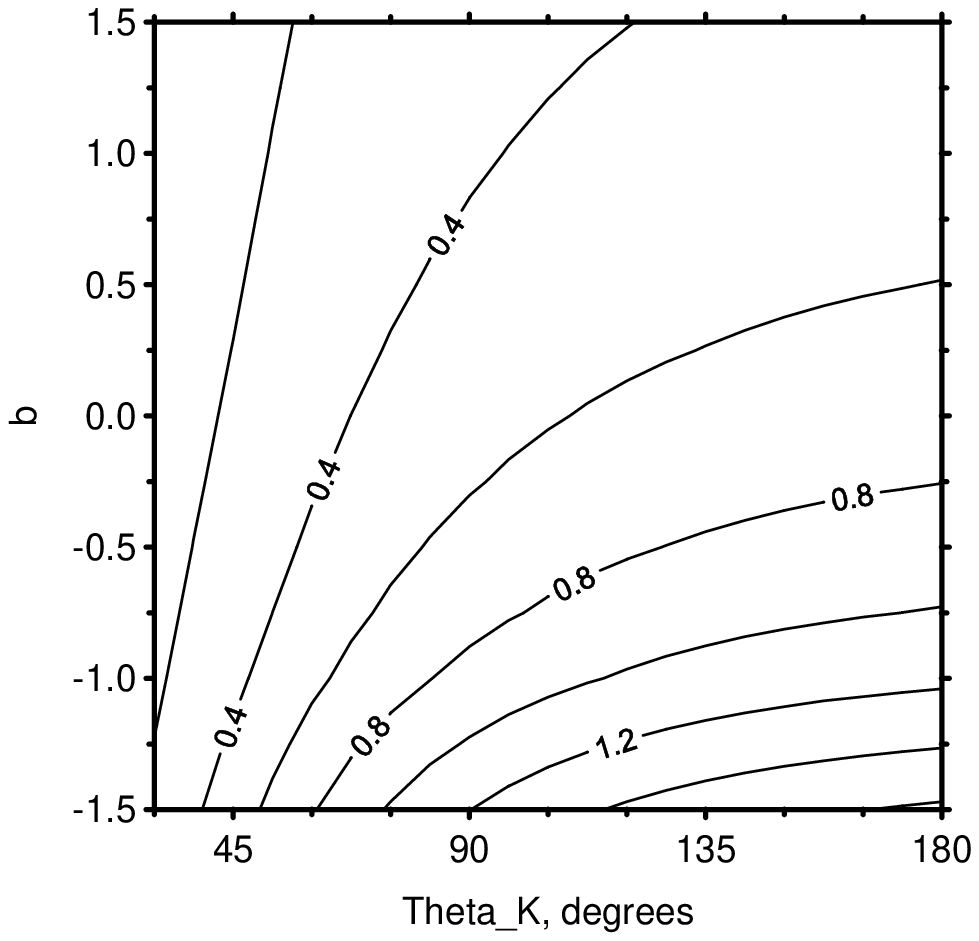} 
\caption{$\zeta\rs{T}$ for different values of parameters $b$ and $\Theta\rs{K}$. $s=2$
              } 
\label{SN1006cp:zetaIC}
\end{figure}
\begin{figure}
\centering
\includegraphics[width=8.3truecm]{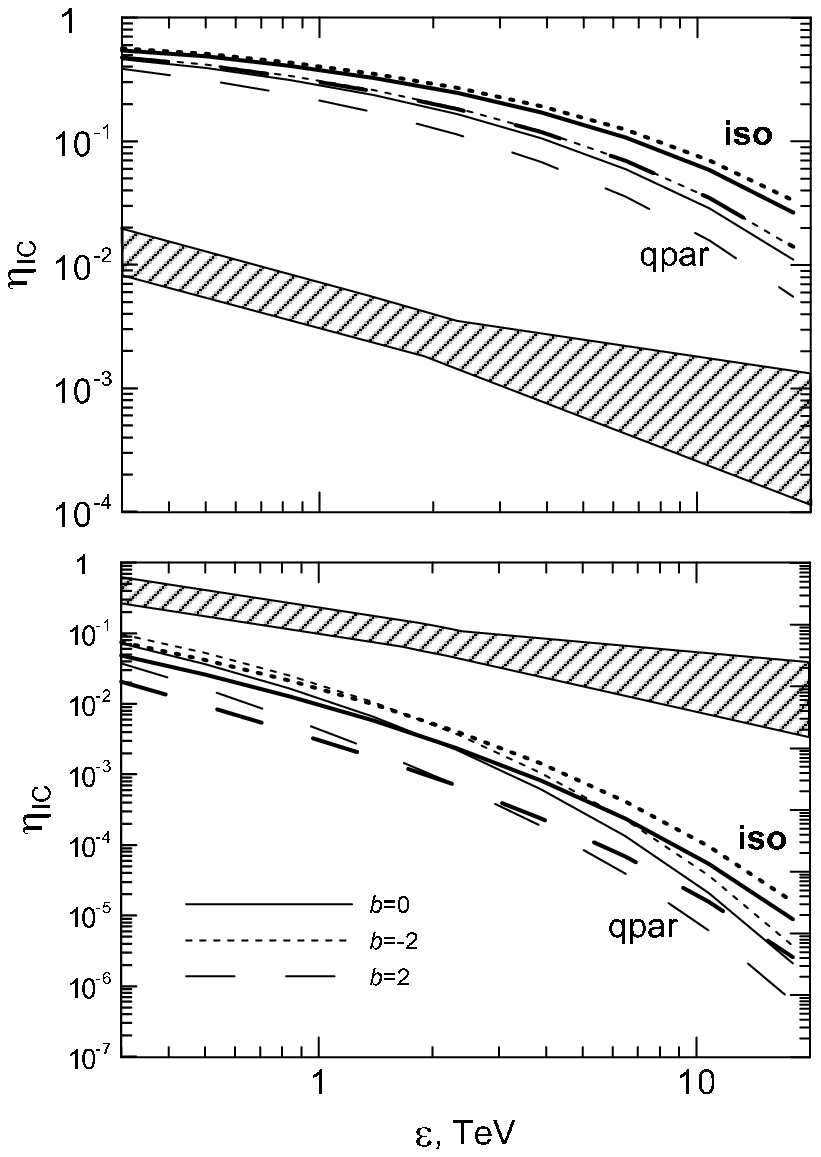} 
\caption{Modification factor $\eta\rs{ic}$. 
Lines are the same as on Fig.~\ref{SN1006cp:etaXfig}. 
Experimental modification factor for SN~1006 are shown for comparison. 
It is obtained from the HESS data \citep[][]{HESS-SN1006-2010} with the use of Eq.~(\ref{sn1006cp:etagobs}) 
and MF strength 
$B\rs{o}=3\un{\mu G}$ (upper panel) and 
$B\rs{o}=30\un{\mu G}$ (lower panel).
              } 
\label{SN1006cp:etaGfig}
\end{figure}

The contribution from electrons with energies around $E\rs{max}$ may be important for TeV \g-photons. The full 
expression for IC process is 
\begin{equation}
 F\rs{ic}(\varepsilon)=C\rs{ic} \zeta\rs{ic}(\varepsilon;b,\Theta\rs{K},\epsilon\rs{f\|},E\rs{max\|})   
                     K\rs{s\|}R^3d^{-2}
 \label{SN1006cp:Fic2}
\end{equation}
where 
\begin{equation}
 C\rs{ic}=\frac{3\sigma\rs{T}kT\rs{CMB}(m\rs{e}c^2)^{3-s}}{2h^3c^2}, 
\end{equation}
\begin{equation}
\begin{array}{l}
 \zeta\rs{ic}({\varepsilon};b,\Theta\rs{K},\epsilon\rs{f\|},E\rs{max\|})=
 \displaystyle
 \int\limits_{0}^{2\pi}\!d\varphi\!
 \int\limits_{0}^{\pi}\!\!d\theta\sin\theta f\rs{K}\!
 \int\limits_{0}^{1}\!\!d\bar a \bar r^{2} \bar r\rs{\bar a} 
 \bar K \! 
 \\ \\ \quad \times\displaystyle
 \int\limits_{\gamma\rs{min}(\varepsilon)}^{\infty}\!d\gamma \gamma^{-2-s}{\cal E}\rs{rad}^{s-2}
 \exp\left(-\frac{\gamma
 }{\gamma\rs{max\|}{\cal E}\rs{ad}{\cal E}\rs{rad}f\rs{E}}\right)
 {\cal I}(\varepsilon,E),
\end{array} \!\!\!\!\!
 \label{eta-f-la3}
\end{equation}
where $\gamma$ is the electron Lorentz factor, 
${\cal I}$ is an integral appearing in the theory of inverse-Compton process \citep[e.g.,][]{Pet08IC}; it accounts for the KN decline where nesessary. 

In case $s=2$ and $T\rs{CMB}=2.75$, ${\cal A}\rs{T}=0.710$ and $C\rs{T}=1.304\E{-14}\un{cgs}$, $C\rs{ic}=1.186\E{12}\un{cgs}$.

In the limit $E\ll E\rs{max}$ and $E\ll \epsilon\rs{f}E\rs{max}$, one has ${\cal E}\rs{rad}=1$ and 
${\cal I}\propto \varepsilon$, $E\rs{min}\propto \varepsilon^{1/2}$ \citep{Pet08IC} 
and (\ref{SN1006cp:Fic2}) transforms to (\ref{SN1006cp:Fic1}). 
Therefore 
\begin{equation}
 \zeta\rs{ic}({\varepsilon})= c\rs{o} 
 {\varepsilon}^{-(s-1)/2}\zeta\rs{T}
 \label{SN1006:eqzeta-ic}
\end{equation}
in this limit; $c\rs{o}=C\rs{T}/C\rs{ic}$.

Let us introduce the modification factor for IC spectrum:
\begin{equation}
 \eta\rs{ic}(\varepsilon,\epsilon\rs{f\|},E\rs{max\|})=\frac{\zeta\rs{ic}({\varepsilon},\epsilon\rs{f\|},E\rs{max\|})
 {\varepsilon}^{(s-1)/2}}
 {c\rs{o}\zeta\rs{T}}.
 \label{modfIC}
\end{equation}
It is also defined to be $\eta\rs{ic}\leq 1$ and ensure $\eta\rs{ic}\rightarrow 1$ well below TeV energies. 
However, it is not so universal as for the synchrotron emission, Eq.~(\ref{modfsynch}): it does not 
scaled with the frequency and it depends on the absolute value of $E\rs{max}$. 
The expression for the broadband IC spectrum is 
\begin{equation}
 F\rs{ic}(\varepsilon)=C\rs{T} \varepsilon^{-(s-1)/2} \zeta\rs{T}(b,\Theta\rs{K}) 
  \eta\rs{ic}(\varepsilon,\epsilon\rs{f\|},E\rs{max\|}) 
                     K\rs{s\|}R^3d^{-2}.
 \label{SN1006cp:Fic5}
\end{equation}

The parameter $\zeta\rs{T}$ behaves like $\zeta$ (Fig.~\ref{SN1006cp:zetaIC}): it mostly depends on $\Theta\rs{K}$ for quasiparallel injection and on $b$ for isotropic injection. However, the role of $\zeta\rs{T}$ is less important for normalization of IC spectrum because it varies in about 4 times over the parameter space. 

The modification factor of the IC specrum $\eta\rs{ic}$ is shown on Fig.~\ref{SN1006cp:etaGfig}, in comparison with the observational data for SN~1006. 
It depends on $\epsilon\rs{f\|}$, $b$, $\Theta\rs{K}$, $s$ and $E\rs{max}$ as well as on the function $f\rs{E}(\Theta\rs{o})$. 

\label{lastpage}
\end{document}